\documentclass[aps,prb,superscriptaddress,floatfix,twocolumn,longbibliography]{revtex4-2}

\usepackage{amsmath,dcolumn,bm,amsthm,tabularx,subfigure}
\usepackage[colorlinks=true,urlcolor=blue,citecolor=blue,linkcolor=blue]{hyperref}
\usepackage{mathrsfs,times}
\usepackage{bbold,dsfont}
\usepackage{graphicx,graphics,color}
\usepackage{mathtools,nicefrac}
\usepackage{slashed}
\usepackage{amsfonts,amssymb}
\usepackage{multirow}
\usepackage{booktabs}
\allowdisplaybreaks

\newcommand{\beq}{\begin{equation}}
\newcommand{\eeq}{\end{equation}}
\newcommand{\bea}{\begin{eqnarray}}
\newcommand{\eea}{\end{eqnarray}}

\newcommand{\R}{{\mathcal{R}}}
\newcommand{\Lft}{{\mathcal{L}}}

\begin{document}

\title{Symmetry- and energy-resolved entanglement dynamics in a disordered Bose-Hubbard Model}

\author{Jie~Chen}
\thanks{These authors contributed equally to this work}
\email[\\Contact author:\ ]{chenjie666@xhu.edu.cn}
\affiliation{Key Laboratory of Artificial Structures and Quantum Control (Ministry of Education), School of Physics and Astronomy, Shanghai Jiao Tong University, Shanghai 200240, China}
\affiliation{School of Science, Key Laboratory of High Performance Scientific Computation, Xihua University, Chengdu 610039, China}

\author{Chun~Chen}
\thanks{These authors contributed equally to this work}
\email[\\Contact author:\ ]{chunchen@sjtu.edu.cn}
\affiliation{Key Laboratory of Artificial Structures and Quantum Control (Ministry of Education), School of Physics and Astronomy, Shanghai Jiao Tong University, Shanghai 200240, China}

\author{Xiaoqun~Wang}
\email[Contact author:\ ]{xiaoqunwang@zju.edu.cn}
\affiliation{School of Physics, Zhejiang University, Hangzhou 310058, Zhejiang, China}
\affiliation{Collaborative Innovation Center of Advanced Microstructures, Nanjing University, Nanjing 210093, China}

\date{\today}

\begin{abstract}

The phenomenology of many-body localization (MBL) develops mainly from tackling the 1D spin or fermion systems. The situation when interacting bosons get clustered in a random potential remains less explored. Using numerical quantum quenches with new and special emphasis on the integration of both symmetry and energy resolutions, we comprehensively study the dynamics of symmetry-resolved entanglement in a disordered Bose-Hubbard (dBH) model, concentrating on the two types of inhomogeneous initial states to target the lower- and higher-energy sections of its dynamical phase diagram. (i) Motivated by the recent experiment [A. Lukin {\it et al}., \href{https://doi.org/10.1126/science.aau0818}{Science {\bf 364}, 256 (2019)}] which focused on the lower-energy dynamical behaviors of the dBH chain, we first show that, at low energies, for a thermalizing state, although the second law of thermodynamics prohibits the decrease of the total entropy over time, for part of the channel-resolved entropies, a long-term entropic reduction may arise at weak disorder. (ii) A companion channel-resolved analysis at strong disorder further hints that the priorly observed double-log growth of the number entropy might not directly indicate the breakdown of MBL in spin or fermion chains, providing a refreshing perspective on this major controversy in the community. (iii) From time-evolving the line-shape low-energy product state, we subsequently reveal an abrupt formation of a novel \lq\lq entropy imbalance pattern'' across the different symmetry channels. Intriguingly, this imbalance melts in the strong-disorder limit. We conjecture that the melting of the entropic pattern, together with the freezing of a concurrent particle-density wave, embodies a dual trait inherent to MBL. (iv) Conversely, the higher-energy section of the dynamical phase diagram is where the dBH model differs most significantly from the spin or fermion systems. This parametric space was not studied in the previous literature. Specifically, we find a cluster MBL regime, unique to the Bose statistics, emerging from the higher-energy section. {\color{black} This cluster MBL regime realizable even at weak disorder does not appear to suffer from the finite-size drift and is distinguished by its absence of the hallmark of MBL---the unbounded growth of the entanglement entropy.} Our theoretical predictions are by and large testable via the present experimental facilities.

\end{abstract}

\maketitle


\section{Introduction}

Arguably, the phenomenology of many-body localization (MBL) might comprise a leading paradigm of the nonergodic eigenstate matter beyond the celebrated Anderson insulator \cite{Anderson,Basko2,Gornyi,Oganesyan,PalPRB,Huse,Abanin,Sierant2024}. The intricate interplay between the randomness and the interaction renders it essential to attack this nonequilibrium problem right from the level of many-body wavefunctions. Since the early days, entanglement entropy and its quantum quench dynamics have been widely deployed to probe the slow albeit unrestricted information propagation in these putative MBL systems \cite{Znidaric,BardarsonPollmannMoore,Serbyn2013,Huang}. Recently, a key advance akin to this reasoning is the imposition of the symmetry resolution. Particularly, the ensuing symmetry-resolved entanglement dynamics has now been successfully measured in the quantum optical lattices using the $^{87}$Rb atoms to witness the logarithmic signature of MBL in the dBH chains \cite{LukinGreiner,leonard2023probing}.

\begin{figure}[t]	
\begin{center}
\includegraphics[width=0.9\linewidth]{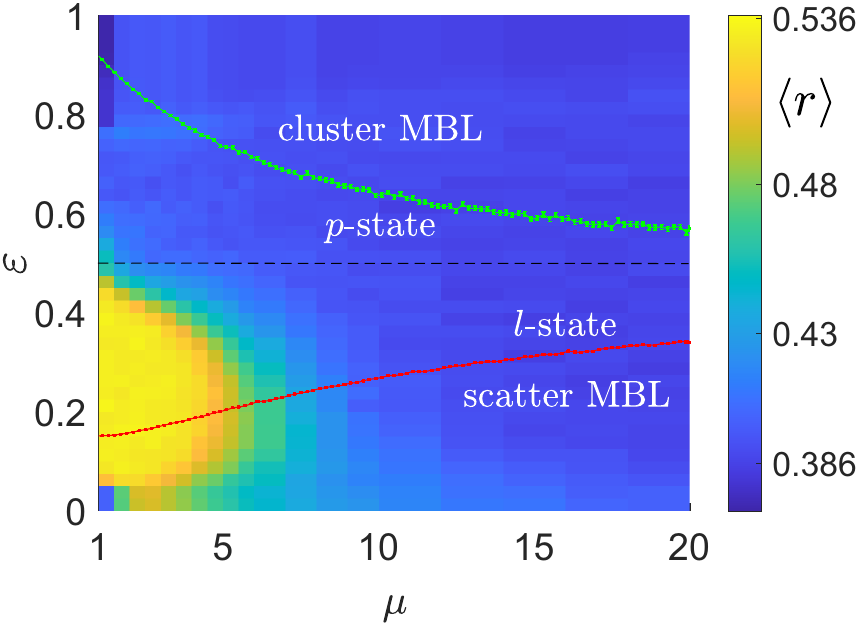}
\caption{Dynamical phase diagram of the periodic dBH model by contour plotting the averaged level-spacing ratio $\langle r\rangle$ \cite{Oganesyan,Atas} of a chain with length $L=12$ and occupation $N=6$. $\varepsilon$ and $\mu$ denote the normalized energy and the disorder strength. To minimize the interaction, the initial $l$-state consists of one (zero) boson on each site of the left (right) half-chain whose normalized energy is traced by the red line. To maximize the interaction, the initial $p$-state accommodates all the bosons on the leftmost site, leaving the remainder unoccupied, whose normalized energy is delineated by the green line. {\color{black} Concretely, for a random sample, we use ED to find all eigenvalues of (\ref{hdbh}) arranged in an ascending order $E_1\leqslant E_2\leqslant\cdots\leqslant E_D$ and $D$ is the dimension of the Hamiltonian matrix. The nearest-neighbor gap ratio $r_n=\textrm{min}(\delta E_n,\delta E_{n-1})/\textrm{max}(\delta E_n,\delta E_{n-1})$ with $\delta E_n=E_n-E_{n-1}$ the gap between the two adjacent eigenenergies of (\ref{hdbh}). The level-spacing ratio $\langle r\rangle$ is obtained from averaging over the $17$ eigenenergies closest to a specified $\varepsilon$ and then over all the $3000$ random samples.}}
\label{pic1}
\end{center}
\end{figure}

Despite being the \lq\lq standard model'' for the interacting bosons, in the context of MBL research, the dBH chain is considerably less explored than its fermionic counterpart, the spinless $t$-$V$ model, or equivalently, the Heisenberg $X\!X\!Z$ chain via the Jordan-Wigner transformation. To our knowledge, even for the existing literature, the dynamical phase diagram of the dBH model has only been worked out occasionally \cite{orell2019probing,Sierant_2018,chen2024eigenstate,Sierant2017}. To this end, one central objective of this article is to construct and elucidate the dynamical phase diagram for the dBH model (see Fig.~\ref{pic1}). On the one hand, as the scatter bosons are the primary low-energy entities or carriers, the lower-lying portion of Fig.~\ref{pic1} resembles the whole phase diagrams of the typical disordered spin or fermion chains \cite{Luitz}. On the other hand, the higher-energy portion of Fig.~\ref{pic1} is unique to the Bose systems because multiple bosons cluster. This binary aspect of the phase diagram, stemming from the particles' quantum statistical properties, demands that the dBH model shall be studied in an energy-resolved manner.

In the present work, we fuse together the aforementioned two key ingredients:
\begin{itemize}
\item the symmetry resolution in the entanglement decomposition,
\item and the energy resolution in the initial-state preparation,
\end{itemize}
as the proper methodology to scrutinize the quantum quench dynamics for the important but still understudied dBH model system.

Specifically, as shown by Fig.~\ref{pic1}, a line-shape initial product state called the $l$-state that minimizes the interaction: 
\begin{equation}
l\textrm{-state}=|1_1,1_2,\ldots,1_{\frac{L}{2}},0_{\frac{L}{2}+1},\ldots,0_L\rangle,
\label{lstate}
\end{equation}
is designed to access the lower-energy section of the phase diagram. {\color{black} Here $1_{i}$ means that there is $1$ boson on the site $i$ and $i=1,...,L/2$. Similarly, $0_j$ means that there is $0$ boson on the site $j$ and $j=L/2+1,...,L$.} Complementarily, a point-shape initial product state called the $p$-state that maximizes the interaction:
\begin{equation}
p\textrm{-state}=|N_1,0_2,\ldots,0_L\rangle,
\label{pstate}
\end{equation}
is devised to assess the opposite higher-energy section. {\color{black} Likewise, here $N_1$ means that there are $N$ bosons on the site $1$, while $0_j$ means that there is $0$ boson on the site $j$ and $j=2,...,L$.} Note that all the bosons are released within the left half-chain, whereby the dominant tendencies of the particle flows in the $l$- and $p$-states are unidirectional toward the right part. This means that particles initially inhabit the largest symmetry-resolved channel, and over time, they migrate toward the smaller channels. Such a setup enables a more oriented exploration of the multi-boson dynamics within the configuration space of entropy.

At low energies, as the disorder strength increases, the system potentially crosses over from a thermalized state to an MBL regime, exhibiting dynamics similar to that in the fermionic or spin models. For the thermalized state at weak disorder, although the fundamental principle of thermodynamics prohibits a decrease of the total entropy over time, we observe a long-term, nontransient depletion in the channel-resolved entropies. For the case of strong disorder, an associated channel-resolved analysis further implies that the formerly observed double-log growth of the number entropy may not indicate the breakdown of MBL for those spin or fermion chains \cite{KieferEmmanouilidisPRL,Luitz2020}. Moreover, in quantum quenches from the $l$-state, an exotic \lq\lq entropic imbalance pattern'' pops up in the coordinate plane of the symmetry index and the time. Unlike the freezing of the particle- or charge-density wave (CDW), this entropy imbalance symmetrizes in the strong-disorder limit. Although the fate of the disorder-induced MBL as an eigenstate phase is currently still under intense debate \cite{Suntajs,Abanin2021,morningstar2022avalanches}, this entropy symmetrization process we propose here relies on the symmetry and the strong-disorder condition but not necessarily on the existence of the MBL phase itself, meaning that this newly found nonequilibrium phenomenon could be sufficiently generic to survive in the thermodynamic limit.

Conversely, at high energies, a cluster MBL regime is found to be stabilized even at the weak-disorder condition, and is distinguished by its absence of the widely assumed unbounded entanglement entropy growth. Further, the cluster MBL regime, unique to the interacting bosons, also features the prolonged inhomogeneities in both the particle and entropy distributions. As the mechanism underpinning the cluster MBL regime need not be the same as that for the MBL regime in the spin or fermion systems, the challenges and critiques from the ongoing debate might not be immediately pertinent to the case at hand, leaving the door potentially open toward harboring the stable MBL phase in Bose systems.

Compared to the previous literature \cite{LukinGreiner,orell2019probing,Sierant_2018,Sierant2017}, the present study has several noteworthy advances. (i) Although the pioneering work \cite{LukinGreiner} launched both the experimental and the theoretical investigations on the symmetry-resolved entanglement dynamics, the primary focus of \cite{LukinGreiner} was however oriented toward the lower-energy section of the dBH chain with the higher-energy section unexplored. Meanwhile, the energy-resolved dynamical phase diagram of the dBH chain at a different filling was previously worked out by \cite{Sierant_2018}, but as it was done before the work \cite{LukinGreiner}, no symmetry resolution in the entanglement measure was discussed. Moreover, \cite{Sierant_2018} also concentrated only on the lower-energy section, leaving the higher-energy section of the phase diagram untouched. In this sense, our study might be the first to treat both the symmetry resolution and the energy resolution on an equally important footing for a prototypical disordered many-body system. (ii) Armed with this new strategy, we further propose the utility of the symmetry-resolved number entropy and the symmetry-resolved entanglement entropy to complement the mainstream approach of \cite{LukinGreiner} which instead is based on the total number and configuration entropies. This new route not only allows us to look into the origin of the double-log growth of the total number entropy deep inside the lower-energy, strong-disorder section of the phase diagram, but it also enables us to unravel both an unusual entropy reduction as well as an exotic entanglement imbalance pattern generated discontinuously from the quantum quench. (iii) Finally, stimulated by the central question regarding the situation when the multiple interacting bosons get clustered in a random potential, we successfully identify the robust and peculiar localization signature to bolster the plausible existence of the cluster MBL regime in the higher-energy section of the dynamical phase diagram of the 1D dBH model. The territory of this cluster MBL regime seems to be largely missing or uncharted in the previous literature. It is important to note that although the density of states within the higher-energy section of the phase diagram is small when the disorder is weak, our previous work \cite{chen2024eigenstate} demonstrated that the successive increase of the disorder strength would appropriately enhance the density of states across the entire higher-energy section. Further, as hinted by our present results, at the weak disorder it is not very possible for the dBH chain to become fully thermalized when the normalized energy is high (see Fig.~\ref{pic1}). The claimed cluster MBL regime in the dynamical phase diagram could therefore avoid the issue of the finite-size drift \cite{ChenChenWang2025} and persist in the asymptotic limit of $L,\ t\rightarrow\infty$. {\color{black} Notably, the nonthermal behavior was reported in the longer clean BH chain \cite{Kollath,Carleo}. It is currently understood that both works \cite{Kollath,Carleo} focused on the lower-energy section of the phase diagram (as the initial states used have a low local occupancy) and the nonchaotic dynamics observed is due to the Hilbert-space fragmentation \cite{SalaPollmann} arising from the large onsite Hubbard repulsion demanded by both works. In contrast, this work focuses on the small onsite Hubbard repulsion. We show in Fig.~\ref{pic9} that the clean BH chain in this case is thermalized in both the lower- and higher-energy sections, demonstrating the irrelevance of the Hilbert-space fragmentation to the cluster MBL regime we find. Recall that the Hilbert-space fragmentation is anticipated to be most pronounced in the disorder-free systems.}

The remainder of the present paper is structured as follows. In Sec.~\ref{sectionmodelsymm}, we introduce the dBH model Hamiltonian, underscore its particle number conservation, and detail the entanglement decomposition formalism based on such a U(1) symmetry. Two separate but complementary routes are then taken to substantiate the bosonic model's entanglement dynamical characteristics. In Sec.~\ref{sectionrltpartone}, we first adopt the more popular strategy by looking into the real-time evolutions of the total number and configuration entropies inside the varied regimes of the dynamical phase diagram, which jointly reflect how the total entanglement entropy would evolve under the tuning of the normalized energy and the disorder strength. In Sec.~\ref{sectionrltparttwo}, the new tactic that explicitly involves both the symmetry and the energy resolutions is proposed and implemented to expose further the hidden inner structure of the entanglement dynamics for the dBH model. Particularly, a robust entanglement imbalance pattern is found to be dynamically engendered from the quantum quench of a product state, whose role under the impact of disorder is then examined and clarified. Finally, Sec.~\ref{sectionsummary} summarizes our major findings on the purely entanglement quench dynamics of the interacting bosons in a random potential and proposes the possible future extensions and directions. {\color{black} In Appendixes~\ref{app1}-\ref{app4}, supporting materials with the data analyses, the explicit mention of the caveat on the instability of the scatter MBL regime, and the preliminary analytics on the number entropy in the lower-energy section of the phase diagram are provided for their relevance to our work.}

\section{Model and symmetry} \label{sectionmodelsymm}

The periodic dBH chain is describable by the following Hamiltonian
\begin{equation}
H_{\textrm{dBH}}=-J\sum^L_{i=1}(a^{\dagger}_ia_{i+1}+\textrm{H.c.})+\sum^L_{i=1}\frac{U}{2}n_i(n_i-1)+\sum^L_{i=1}\mu_in_i
\label{hdbh}
\end{equation}
where $a^\dagger_i\ (a_i)$ is the boson creation (annihilation) operator at site $i$, $n_i=a^\dagger_ia_i\ (N=\sum^L_in_i)$ counts the local (total) boson occupation number, $U$ parametrizes the onsite Hubbard interaction, and $\mu_i\in[-\mu,\mu]$ is a diagonal random potential drawn from the box distribution. Importantly, $[N,H_{\textrm{dBH}}]=0$, so the number-conserving dBH model respects the U(1) symmetry. In this work, all the relevant quantities are averages over at least 1000 random samples, solved by exact diagonalization (ED) \cite{Zhang2010} or the Krylov-iterative method \cite{Paeckel}. We set $J=1$ as the energy unit and fix $U=3J,\ N=\frac{L}{2}$ in the succeeding numerical calculations.

\begin{figure*}[htb]
\begin{center}
\includegraphics[width=1\linewidth]{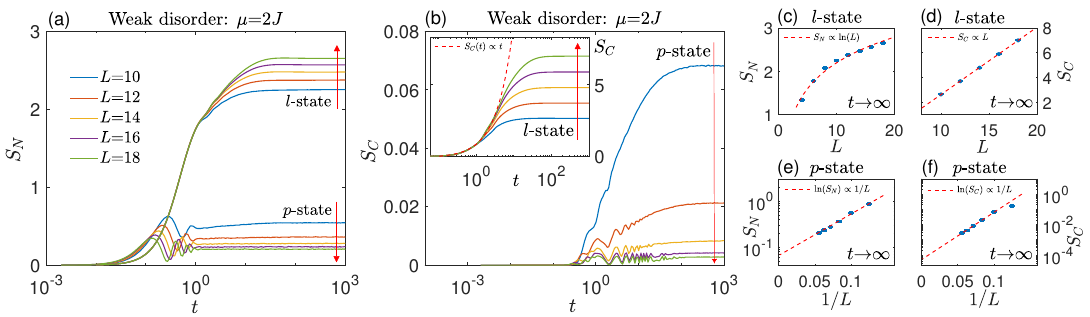}
\caption{Quantum quench dynamics of the half-chain (a) $S_N$ and (b) $S_C$ for the dBH model, focusing on the weak-disorder section $(\mu=2J)$ where the ETH and the cluster MBL regimes (see Fig.~\ref{pic1}) are realizable by commencing from the $l$- and $p$-states, respectively. To evolve the longer chains $(L=16,18;\ N=\frac{L}{2})$, the Krylov method \cite{Paeckel} is employed. (c)-(f) highlight the corresponding scaling behaviors of the saturation values of $S_N$ and $S_C$ for the $l$- and $p$-states as a function of $L$ and $1/L$. {\color{black} See Appendix~\ref{app4}(i),(ii) for the data analyses.}}
\label{pic2}
\end{center}
\end{figure*}

Denote the total conserving operator as $Q=Q_\Lft+Q_\R$ which is separable into two parts $\Lft$ and $\R$, then for an eigenstate $|\psi\rangle$ of $Q$, the reduced density matrix of $\Lft$, $\rho_\Lft={\textrm{Tr}}_\R(|\psi\rangle\langle\psi|)$, commutes with $Q_\Lft$, implying $\rho_\Lft=\oplus_n\rho_{\Lft,n}$ where $\rho_{\Lft,n}$ is the assembly of the blocks possessing the same eigenvalue $n$ of $Q_\Lft$. Because $\sum_n{\textrm{Tr}}_{\Lft,n}\rho_{\Lft,n}=1$, then for $|\psi\rangle$, $p_n={\textrm{Tr}}_{\Lft,n}\rho_{\Lft,n}$ represents the probability of yielding the eigenvalue $n$ in the projective measurement of $Q_\Lft$. Within that subspace, the normalized reduced density matrix assumes $\tilde{\rho}_{\Lft,n}=p^{-1}_n\rho_{\Lft,n}$. Consequently, the entanglement entropy of $\Lft$ decomposes into \cite{donnelly2012decomposition}
\begin{equation}
S_{\textrm{vN}}=-\sum_np_n\log_2p_n+\sum_np_nS^n_{\textrm{vN}}
\label{totalsvn}
\end{equation}
where
\begin{equation}
S^n_{\textrm{vN}}=-{\textrm{Tr}}_{\Lft,n}\tilde{\rho}_{\Lft,n}\log_2\tilde{\rho}_{\Lft,n}
\label{svnn}
\end{equation}
becomes the symmetry-resolved entanglement entropy \cite{Goldstein,Banerjee2024,CastroAlvaredo,turkeshi2020entanglement,Murciano2020,ares2023entanglement} for $\Lft$ accommodating $n$ bosons.
Most previous works focus on the so-called number and configuration entropies, 
\begin{gather}
S_N=-\sum_np_n\log_2p_n, \\
S_C=\sum_np_nS^n_{\textrm{vN}},
\end{gather}
which quantify the respective total entropies from the particle or charge fluctuations across different sectors and the configurational superpositions within each sector weighted by the probability \cite{Goldstein,Xavier,LukinGreiner,Bonsignori,ParezPRB,KieferEmmanouilidisPRL,Luitz2020,KieferEmmanouilidisPRB,Feldman,Jeyaretnam2024,CastroAlvaredo,Banerjee2024,singh2016signatures}.

The subsequent results exploiting the above formalism are primarily partitioned into two parts. In the first part (Sec.~\ref{sectionrltpartone}), we systematically explore the overall dynamics of the number entropy $S_N$ and the configuration entropy $S_C$ across the various phase regimes. In the second part (Sec.~\ref{sectionrltparttwo}), we delve into the internal structure of the entanglement entropy $S_{\textrm{vN}}$, taking advantage of the set of mutually independent entanglement measures $\{n, p_n, S^n_{\textrm{vN}}\}$, where $p_n$ represents the absolute weight while $S^n_{\textrm{vN}}$ is determined by the relative weight \cite{ChenChenWang2024}. In terms of the channel or sector index $n$, one can dynamically contrast $S^{n}_{\textrm{vN}}$ where $p_{n}$ a maximum with $S^{N-n}_{\textrm{vN}}$ where $p_{N-n}$ a minimum to uncover the purely entanglement structures beyond the usual scheme of space and time. 

It is worth stressing that the time evolution of $p_n$ itself as well as the time evolution of a correlator
\begin{equation}
C_n\coloneqq\sum_{\{\Lft_n\}}\sum_{\{\R_{N-n}\}}|p(\Lft_n\otimes \R_{N-n})-p(\Lft_n)p(\R_{N-n})|
\end{equation}
that is complementary to $S^n_{\textrm{vN}}$ were both measured in the real experiments \cite{LukinGreiner,leonard2023probing,kranzl2023observation}. Here, $\{\Lft_n\}~(\{\R_{N-n}\})$ denotes all the possible configurations with $n~(N-n)$ particles in the left (right) half-chain. $C_n$ is not exactly $S^n_{\textrm{vN}}$, but \cite{LukinGreiner} suggests that it might capture the qualitative characteristics of $S^n_{\textrm{vN}}$ through the quantification of the separability between $\Lft$ and $\R$. In this regard, our predictions below may largely be observable.

\section{Route one: The entanglement dynamics in terms of the total \boldmath{$S_N$} and \boldmath{$S_C$}} \label{sectionrltpartone}

To set the stage, in this section we explore the quantum quench dynamics of the entanglement entropy $S_{\textrm{vN}}$ for the dBH model via the examination upon the time evolutions of its two component parts, i.e., the total number entropy $S_N$ and the total configuration entropy $S_C$, as a function of the initial-state configuration and the model's disorder strength. To understand the several peculiarities arising from the obtained results, we then introduce the channel-resolved number entropy and demonstrate that, unlike the total $S_N$, a subset of the channel-resolved number entropies can decrease over time. Moreover, within this channel-resolved framework, the puzzling double-log growth of $S_N$ observed in the lower-energy, strong-disorder section of the dynamical phase diagram might also be interpreted.

\subsection{Dynamical distinctions from \boldmath{$S_N$ \& $S_C$}, and the cluster MBL regime without the unbounded entropy growth}

\begin{figure*}[htb]
\begin{center}
\includegraphics[scale=1.0]{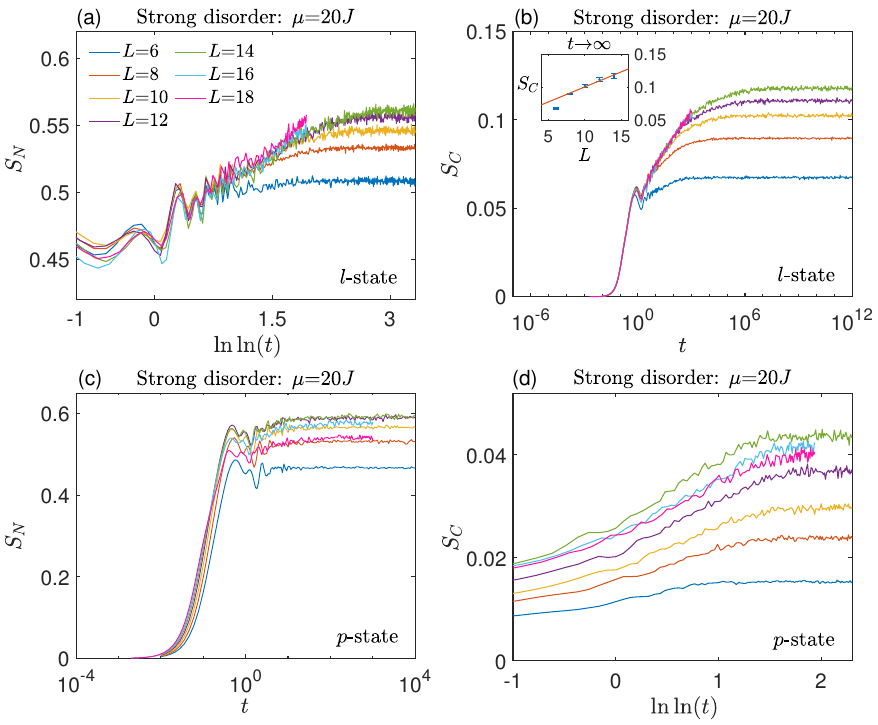}
\caption{Quantum quench dynamics of the half-chain entanglement for the dBH model at strong disorder. The first (second) row is devoted to the scatter (cluster) MBL regime at $\mu=20J$. (a) for $S_N$ and (b) for $S_C$ starting from the $l$-state. (c) for $S_N$ and (d) for $S_C$ starting from the $p$-state. To evolve the longer chains $(L=16,\ 18;\ N=\frac{L}{2})$, again the Krylov method \cite{Paeckel} is used. {\color{black} See Appendix~\ref{app4}(iii)-(v) for the data analyses.}}
\label{pic2_1}
\end{center}
\end{figure*}

We first investigate the dynamics of total $S_N$ and $S_C$, starting from the respective $l$- and $p$-states, under the scenarios of weak and strong disorder. Figures~\ref{pic2} and \ref{pic2_1} summarize the characteristic results of the four types of quantum quench dynamics perceived from this perspective.

Figures~\ref{pic2}(a)-(f) target the weak-disorder section of small $\mu$. For the thermalization dome in the dynamical phase diagram (Fig.~\ref{pic1}) accessible via the lower-energy $l$-state, besides reproducing the known linear-$t$ and linear-$L$ scalings for the growth and saturation of $S_C$ [inset of Fig.~\ref{pic2}(b) and Fig.~\ref{pic2}(d)], we find that the temporal buildup of $S_N$ obeys a logarithmic function of $t$ \cite{Maximilian2020,kiefer2021slow} and its saturation scales as a logarithmic function of $L$ [see Figs.~\ref{pic2}(a),(c) and Eq.~(\ref{SNlogL})]. By contrast, once switching to the higher-energy $p$-state, both the total $S_N$ and $S_C$ evolutions become halted [see Figs.~\ref{pic2}(a),(b)] and the scalings of their saturations fulfill the area law [see Figs.~\ref{pic2}(e),(f)]. The cluster MBL regime at small $\mu$ hence features a bounded total $S_{\textrm{vN}}$ growth, and an interaction-facilitated entropy inhomogeneity to be quantified shortly. Notably, this latter feature never occurs in the Anderson insulator.

\begin{figure*}[htb]
\begin{center}
\includegraphics[scale=1.0]{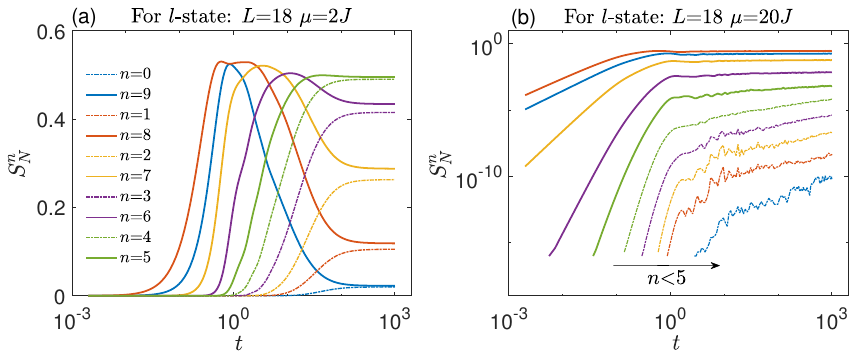}
\caption{Time evolution of the channel-resolved number entropy $S^n_N$. Here we start from the initial $l$-state in a disordered, periodic BH chain of length $L=18$. (a) gives the results for the thermal phase realized at the weak-disorder regime of $\mu=2J$. (b) corresponds to the regime deep inside the scatter MBL regime stabilized by strong disorder $(\mu=20J)$. Notice that the time when $S^n_N$ begins to deviate from zero is delayed with the decreasing $n$ in (a),(b).}
\label{pic3}
\end{center}
\end{figure*}

Figures~\ref{pic2_1}(a)-(d) present the $S_N$ and $S_C$ results for the opposite strong-disorder section. Figure~\ref{pic2_1}(a) shows that the total $S_N$ growth in the scatter MBL regime, by starting from the $l$-state, follows a double-logarithmic function of $t$ at large $\mu$ \cite{KieferEmmanouilidisPRL,sierant2022challenges}. Subsequently, via a channel-resolved analysis, we show in Fig.~\ref{pic3} that this perceived particle-number fluctuation occurs mainly as the reorganization within the initial number channels where the particles are first released rather than the substantial particle transport involving all the available channels, in particular, those remote ones. To wit, the scaling of the $S_N$ saturation remains obedient to the area law. This is in accord with Fig.~\ref{pic3}(b) but differs from the main claim in \cite{KieferEmmanouilidisPRL,KieferEmmanouilidisPRB}. The companion $\ln(t)$ rise of $S_C$ in the scatter MBL regime, along with the volume-law scaling of its saturation, is revealed by Fig.~\ref{pic2_1}(b). Interestingly, the cluster MBL regime at large $\mu$, by starting from the $p$-state, exhibits no appreciable temporal growth in $S_N$, whose saturation thereby obeys an area scaling law as evidenced by Fig.~\ref{pic2_1}(c). The accompanying $S_C$ in the cluster MBL regime, however, grows as a tentatively double-logarithmic function of $t$, but likely saturates to the area law as well at the long-time limit [see Fig.~\ref{pic2_1}(d)].

\def\arraystretch{1.3}
\begin{table}[b]
\caption{Dynamical $S_N$ and $S_C$ characteristics of the dBH model.} 
\label{table1}
\centering
\begin{tabular}{|c|c|cccc|} 
\hline
\multicolumn{2}{|c|}{\multirow{3}{*}{regime}} & \multicolumn{1}{c|}{\multirow{3}{*}{\shortstack{ETH \\ small $\mu$}}} & \multicolumn{3}{c|}{MBL}    \\
\cline{4-6}  
\multicolumn{2}{|c|}{}                       & \multicolumn{1}{c|}{}                     & \multicolumn{1}{c|}{\multirow{2}{*}{\shortstack{scatter \\ large $\mu$}}} & \multicolumn{2}{c|}{cluster}                 \\
\cline{5-6}
\multicolumn{2}{|c|}{}                       & \multicolumn{1}{c|}{}                     & \multicolumn{1}{c|}{}                         & \multicolumn{1}{c|}{small $\mu$} & large $\mu$             \\ 
\hline
\multirow{2}{*}{$S_N$} & growth              & $\ln t$                                     & $\ln\ln t$                                       & \textrm{no}               & \textrm{no}      \\ 
\cline{2-2}
& scaling                 & $\ln L$                                     & \textrm{area}                                 & \textrm{area}             & \textrm{area}   \\ 
\cline{1-2}
\multirow{2}{*}{$S_C$} & growth              & $t$                           & $\ln t$                                         & \textrm{no}               & $\ln\ln t$          \\ 
\cline{2-2}
& scaling                 & \textrm{volume}                           & \textrm{volume}                               & \textrm{area}             & \textrm{area}  \\
\hline
\end{tabular}
\end{table}

Table~\ref{table1} recaps the salient features of the total $S_N$ and $S_C$ evolutions to help differentiate between the four dynamical regimes in the dBH chain. The first two columns of the table present the low-energy behaviors starting from the $l$-state, while the last two columns present the high-energy behaviors starting from the $p$-state. It is noticeable therein that the low-energy behaviors of the dBH model resemble those of the fermionic (spin) models \cite{BardarsonPollmannMoore,Serbyn2013,KieferEmmanouilidisPRL}. In some detail, within the low-energy section accommodating the scatter MBL regime, the unbounded growth of the total entanglement entropy $S_{\textrm{vN}}$ arises from the contribution of the configuration entropy $S_C$, while the part of the number entropy $S_N$ saturates to follow an area law, consistent with the experimental findings in \cite{LukinGreiner}. In the high-energy section accommodating the cluster MBL regime, however, both $S_N$ and $S_C$ saturate to exhibit the area law irrespective of the value of the disorder strength $\mu$. This unusual observation inside the cluster MBL regime challenges the common belief that MBL should always have an unbounded growth upon the measure of the total entanglement entropy $S_{\textrm{vN}}$ \cite{BardarsonPollmannMoore,Serbyn2013,Huang}. The finding of this bounded $S_{\textrm{vN}}$ growth across the cluster MBL regime hence comprises one of the central results of the present work.

\subsection{Channel-resolved number entropy reduction in thermalization}

The total number entropy $S_N$ is defined by $S_N(t)=-\sum_n p_n(t)\log_2 p_n(t)$. This form naturally suggests the parsing of $S_N$ into the channel-resolved number entropy given by
\begin{equation}
S^n_N(t)=-p_n(t)\log_2 p_n(t).
\end{equation}
The second law of thermodynamics dictates that the entropy of a thermalizing state always increases with time \cite{reichl2016modern}. As illustrated by Fig.~\ref{pic2}(a), the total number entropy $S_N(t)$ of the thermalizing state (starting from the $l$-state under $\mu=2J$) indeed increases over time. However, concerning $S^n_N(t)$, as depicted by Fig.~\ref{pic3}(a), we instead notice the stabilization of a long-term decreasing trend over time occurring in those large-$n$ channels. As can be seen from the graph, this decrease extends from $Jt=1$ to $100$, indicating its potential detectability in the actual experiments.

Initially, at the start of the process, all the particles are located on the left side, specifically within the channel $n=9$. As the evolution progresses, particles predominately move toward the right half, leading to a rapid increase of the number entropy in channels $n=9,8,7,6$, and $5$. Additionally, under the condition of low disorder, the system follows eigenstate thermalization hypothesis (ETH) and maintains the approximate left-right symmetry in the long-time steady state. Thus, the steady state respects the channel reflection symmetry, indicating that the particle number entropies in channels $n$ and $N-n$ are roughly equal. This constraint of symmetry results in a subsequent decrease of the channel-resolved number entropies in channels $n=9,\ 8,\ 7,\ 6,\ 5$, and this decrease persists for a significant duration. To our knowledge, such a prolonged decrease of the channel-resolved entropy triggered by an initial-state preparation has not been observed before in the quantum quench dynamics of a thermalizing state.

The total entropy of a closed system never declines. Nonetheless, it is evident that each channel of the isolated system behaves like an open subsystem, mutually coupled to all other channels, whereby the entropy of a certain open subsystem can decrease during the evolution.

\subsection{The double-log growth of \boldmath{$S_N$} in the lower-energy, strong-disorder section arises from localization}

Ref.~\cite{KieferEmmanouilidisPRL} reported that the total number entropy $S_N$ seems to grow double logarithmically over time even deep inside the MBL regime, hinting that the full localization might be unstable in the thermodynamic limit due to the unceasing particle or energy transport.

In this subsection, we perform a channel-resolved inspection of the total number entropy $S_N$ for the dBH chain to help clarify the possibility that the observed double-log growth of $S_N$ might not necessarily indicate the breakdown of the full localization.

\begin{figure*}[htb]
\begin{center}
\includegraphics[width=1\linewidth]{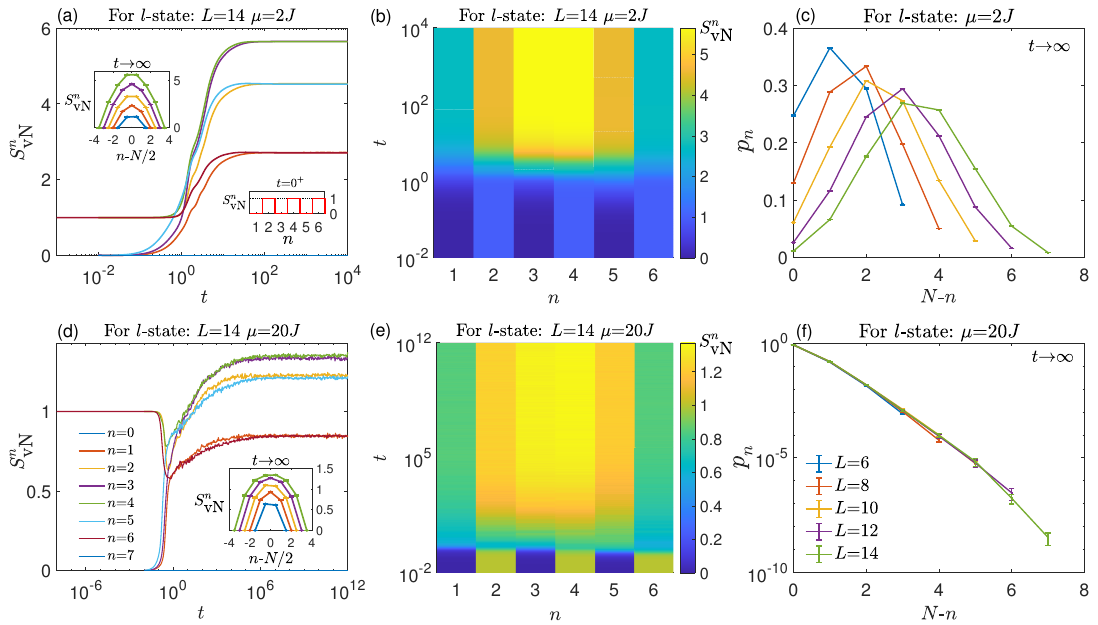}
\caption{Time evolution of the half-chain entanglement entropy from the initial $l$-state resolved into each symmetry channel labelled by the number index $n$. A periodic chain of length $L$ and filling $N=\frac{L}{2}$ is used. The top and bottom rows address the weak- and strong-disorder conditions, while the left, middle, and right columns target $S^n_{\textrm{vN}}(t)$ and the scaling of $p_{n,\infty}$ with $L$. The middle column [(b),(e)] displays the contour plots of $S^n_{\textrm{vN}}(t)$ in the $(n,t)$ plane where the entanglement imbalance pattern [see (a) lower inset] and its melting are demonstrated. Notice that the upper inset of (a) and the lower inset of (d) present the scaling of the saturation values of $S^n_{\textrm{vN}}$ at the infinite time as an increasing function of $L$ (the $n=0,\frac{L}{2}$ components vanish identically). For illustration, the error bars are preserved only in the saturation results but omitted otherwise in the time evolutions.}
\label{pic4}
\end{center}
\end{figure*}

Figure~\ref{pic3}(b) shows the quench evolutions of $S^n_N(t)$ in a log-log format starting from the initial $l$-state for a strongly disordered $(\mu=20J)$, periodic BH chain of length $L=18$. There are two salient features from Fig.~\ref{pic3}(b). 
\begin{enumerate}
\item The large-$t$ temporal growth of $S^n_N$ in those small $n<5$ channels fulfills a power law, which is much faster than $\log\log(t)$. While, for those large $n>5$ channels, the $S^n_N$ growth becomes noticeably slower.
\item The absolute values of $S^n_N$ in those small $n<5$ channels are negligibly smaller than the absolute values of $S^n_N$ in those large $n>5$ channels. Particularly, the contributions from channels $n=9,\ 8$ appear to dominate the whole time evolution of the total $S_N$.
\end{enumerate}

From Fig.~\ref{pic2_1}(a), we already know that the total $S_N$ in this quench setup indeed grows double logarithmically over time. Therefore, combine the above-listed two observations, it is tempting to argue that the double-log growth of $S_N$ shall be predominantly controlled by the channels of large $n=9,\ 8$ and the participation of those small-$n$ channels may be negligible. In other words, in this quantum quench evolution, particles are well confined to the left half-chain. The perceived particle-number fluctuations occur mainly as the reorganizations within the initial number channels where the particles are first released rather than the substantial particle transport involving all the available channels, especially, those remote channels of small $n$. Through this channel-resolved analysis, it is potentially elucidated that the double-log growth of $S_N$ observed in the lower-energy section of the dynamical phase diagram under the strong-disorder condition might largely stem from the fact that the system has been undergoing or approaching localization \cite{Luitz2020,ghosh2022resonance,aceituno2024ultraslow}.

However, see also Appendixes~\ref{app1} and \ref{app2} for the explicit emphasis on the caveat regarding the probable instability of the scatter MBL regime in the asymptotic limit. If the finite-size drift ultimately turned the scatter MBL regime into the thermal phase when $L,\ t\rightarrow\infty$, then to a large extent, the double-log growth of $S_N$ would disappear and Fig.~\ref{pic3}(b) would give way to Fig.~\ref{pic3}(a). In this sense, we argue that according to our channel-resolved analysis, the observed double-log growth of $S_N$ on a finite chain appears more compatible with the phenomenon of localization. 

\section{Route two: The inner structure of the entanglement dynamics via \boldmath{$\{p_n\}$} and \boldmath{$\{S^n_{\textrm{vN}}\}$}} \label{sectionrltparttwo}

In this section, we switch gears to take advantage of the alternative route two to unravel and monitor the internal fine structure of the entanglement dynamics for the dBH model. To this aim, we introduce and exploit the new sets of the mutually independent entanglement measures: $\{p_n\}$ and $\{S^n_{\textrm{vN}}\}$ (see Sec.~\ref{sectionmodelsymm}), which explicitly carry the symmetry index $n$. Incorporate such a tool within the quantum quench protocol allows us to find a novel entanglement imbalance pattern popping up dynamically from a plain initial product-state configuration. The fate of this entanglement pattern, along with how the concurrent particle-density distribution of the initial product state evolves, under the influence of weak and strong disorder constitutes a useful means to help distinguish between the various types of the phase regimes in the dynamical phase diagram of the dBH model.    

\subsection{Dynamically create the entanglement pattern in the lower-energy section}

Most quantum quench studies of MBL start from the nonentangled product states with the predesigned local density imbalance imprinted \cite{SchreiberBloch,BardarsonPollmannMoore,Luitz2016}. As entanglement is absent from the start and usually builds up transiently in a continuous fashion, this construction appears to be structureless in the initial preparation or, more precisely, the initial generation of the entanglement.

Curiously, can the structural features or patterns of the entanglement evolve discontinuously from the product state at an infinitesimal lapse of time?

Intriguingly, the symmetry-resolved entanglement entropy $S^{n}_{\textrm{vN}}$ defined in Eq.~(\ref{svnn}) constitutes an ideal apparatus to address this above question. Figures~\ref{pic4},~\ref{pics1} and \ref{pic5} below show collectively the entanglement growth resolved into each number sector or channel in the numerical quantum quench experiment. Depending on how the bosons are initially populated, two distinct dynamical patterns, one trivial corresponding to the $p$-state and the other nontrivial corresponding to the $l$-state, are observed.

Starting from the $l$-state where the scatter bosons are the leading mobile identity, one novelty of the quantum quench dynamics of the entanglement is the discontinuous jump of $S^n_{\textrm{vN}}$ from $0$ to $1$ at $t=0^+$. As demonstrated by Figs.~\ref{pic4}(a),(d) and \ref{pics1}, for the periodic even chain with the half-filled number of bosons, all the nontrivial $S^{n}_{\textrm{vN}}$ jump to $1$ if $n$ share the same parity of $N-1$. For the other $n$ of the opposite parity of $N-1$, $S^{n}_{\textrm{vN}}$ instead develop smoothly from $0$ up to the saturation. In analog to the familiar CDW order with the local particle number imbalance amongst the even and odd lattice sites, based on the product $l$-state, there arises a \lq\lq nonlocal entanglement imbalance pattern'' amongst the symmetry channels $n$ of the alternating parities, coined the \lq\lq entanglement channel wave (ECW).'' Concretely, the lower insets of Figs.~\ref{pic4}(a),~\ref{pics1}(a)~and~\ref{pics1}(c) illustrate how these resulting $\mathbb{Z}_2$ entanglement patterns look like in a pictorial way.

\subsection{Entropy symmetrization across the lower-energy section}

One significance of the above finding pertains to scrutinizing how it responds to the influence of disorder. Will this entanglement imbalance pattern freeze in the localized regimes like CDW? To set the stage, we first examine what will happen in the thermal phase. As exemplified by Figs.~\ref{pic4}(a)-(c), it turns out that in the weak-disorder section accessible via the initial $l$-state, both the entanglement imbalance pattern and the particle density pattern melt to conform with ETH \cite{Deutsch,Srednicki,DAlessio,KimPRE}. This is because under weak disorder, the reflection symmetry of the clean BH model is broken through a smooth manner, then for each eigenstate within the thermalization energy window, it follows that $S^{N-n}_{\textrm{vN},\Lft}\approx S^{N-n}_{\textrm{vN},\R}$. Next, by virtue of $S^{N-n}_{\textrm{vN},\R}=S^{n}_{\textrm{vN},\Lft}$, valid for arbitrary pure states, one derives that $S^{N-n}_{\textrm{vN},\Lft}\approx S^{n}_{\textrm{vN},\Lft}$, indicating that the early-time even/odd-$n$ entanglement imbalance disappears at the long-time limit. Parallel rationale carries over to the infinite-$t$ profile of $p_n$ (denoted as $p_{n,\infty}$): thermalization dictates that the initially inhomogeneous boson population melts into the final uniform density landscape captured by a Gaussian.

Interestingly, we find that for the initial $l$-state, this entanglement imbalance pattern symmetrizes with respect to the channels $n$ versus $N-n$ even when subject to the strong-disorder condition [see Figs.~\ref{pic4}(d),(e)], suggesting that in the putative scatter MBL regime, the nonlocal entanglement imbalance pattern melts. Concurrently, the companion local boson occupations $p_{n,\infty}$ remain rigidly frozen onto the initially asymmetric form [see Fig.~\ref{pic4}(f)], in accordance with the phenomenology of the full localization on a finite chain. (See Appendixes~\ref{app1} and \ref{app2} for the related caveat on the instability of the scatter MBL regime.)

Assuming the applicability of the local-integrals-of-motion phenomenology for MBL \cite{Serbyn,HuseLIOM,Ros,Anushya,geraedts2017emergent}, then the system's eigenstates at large $\mu$ might be prescribable by filling the localized bits (l-bits). The diagonal approximation after taking the infinite-time limit then informs that the quasi-exponential decay of $p_{n,\infty}$ as a function of $n$ shown in Fig.~\ref{pic4}(f) is determined by the projection coefficients of the initial $l$-state into these l-bit eigenstates. Via interpreting these expansion coefficients as the tunneling amplitudes, it is comprehensible for a finite-length chain that the probability of the corresponding multi-boson tunneling processes is exponentially suppressed in the localized regime as per a measure set by the localization length.

\begin{figure*}[htb]
\includegraphics[scale=0.9]{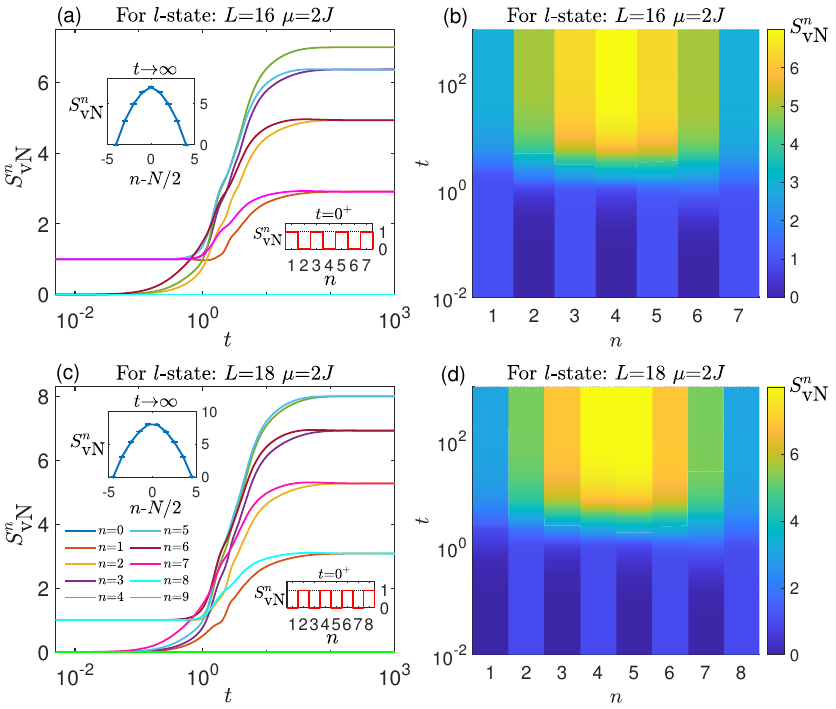}
\caption{Two types of the entanglement imbalance patterns and their different melting processes in the thermal regime of the phase diagram. (a),(b) show the formation and the time evolution of the channel-reflection-symmetric pattern in a periodic even chain accommodating even number of the total bosons. The melting of this first type of the entanglement imbalance pattern removes the $\mathbb{Z}_2$ channel wave pattern but preserves the channel reflection symmetry. (c),(d) show the formation and the time evolution of the channel-reflection-asymmetric pattern in a periodic even chain accommodating odd number of the total bosons. The melting of this second type of the entanglement imbalance pattern not only removes the $\mathbb{Z}_2$ channel wave pattern but also allows for the emergence of the channel reflection symmetry.}
\label{pics1}
\end{figure*}

\begin{figure*}[htb]
\begin{center}
\includegraphics[width=1\linewidth]{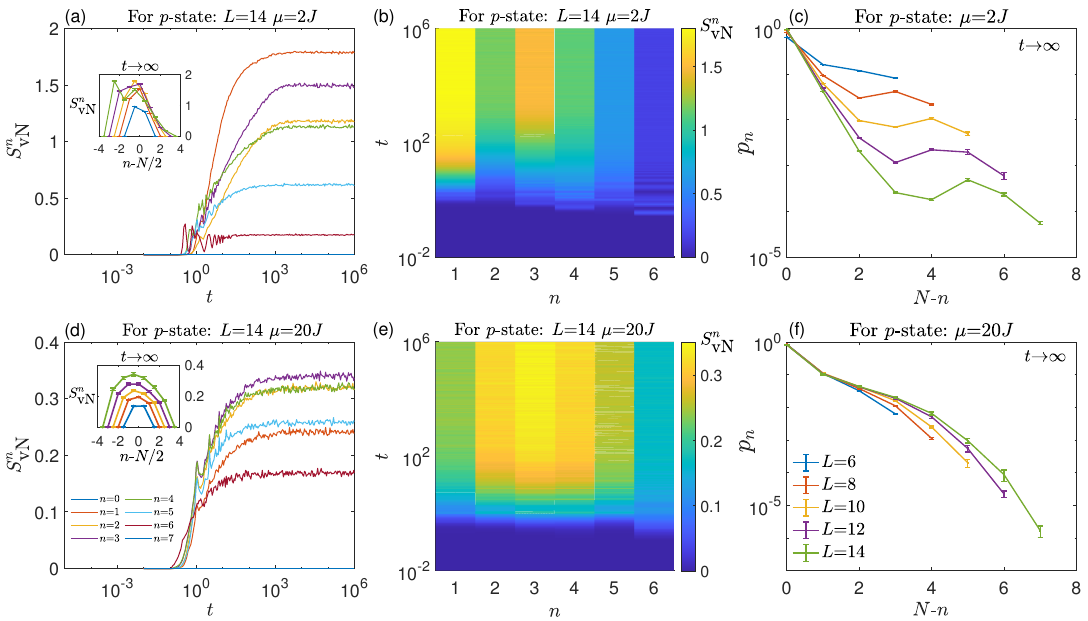}
\caption{Time evolution of the half-chain entanglement entropy from the initial $p$-state resolved into each symmetry channel labelled by the number index $n$. It is worth stressing that the upper insets of panels (a),(d) showcase the scaling of the saturation values of $S^n_{\textrm{vN}}$ at the infinite time as an increasing function of $L$. Other arrangements parallel that of Fig.~\ref{pic4}.}
\label{pic5}
\end{center}
\end{figure*}

One salient feature of Fig.~\ref{pic4} is the coexistence of the resemblance of $\{S^n_{\textrm{vN}}\}$ between (a),(d) with the contrast of $\{p_{n,\infty}\}$ between (c),(f). A phenomenological argument for this disparate trend on a finite-length chain might run as follows. Take the pair of channels $n=1,\ 6$ in an $L=14,\ N=7$ chain as an example, then for a single sample, the use of the normalized $\tilde{\rho}_{\Lft,n}$ makes it possible to examine the relative arrangements of the component states within and between each individual channel of the pair. The overall discrepancy in the prefactors between the two is hidden. As Fig.~\ref{pic4}(f) hints that the finite system might be localized at $\mu=20J$, the dimensions of the pertinent nonzero blocks in $\tilde{\rho}_{\Lft,n=1}$ and $\tilde{\rho}_{\Lft,n=6}$ are controlled by the localization length. This constraint on the multi-boson configurations in each channel, combined with the minimization of the resultant energy mismatch, implies that these dominant density-matrix blocks might largely be diagonal. However, in view of the fact that
\begin{enumerate}
\item[(i)] the $n=1$ channel is dominated by the configurations with $3$ bosons concentrated near the right entanglement cut and another $3$ bosons near the left entanglement cut while the remaining $1$ boson localized at the midpoint of the left half-chain; 
\item[(ii)] the $n=6$ channel is dominated by the configurations with $6$ bosons evenly distributed along the left half-chain and $1$ boson localized at either the left or the right cut;
\item[(iii)] the onsite potentials $\mu_i$ now represent the large and uncorrelated random numbers,
\end{enumerate}
it is not guaranteed that within a single sample, the equality of the saturation values between $S^{n=1}_{\textrm{vN}}$ and $S^{n=6}_{\textrm{vN}}$ ensues. The observed entropy symmetrization in Figs.~\ref{pic4}(d),(e) thus hints that only after averaging over a sufficient amount of the random sample realizations, the statistical distributions of the eigenspectra of $\tilde{\rho}_{\Lft,n=1}$ and $\tilde{\rho}_{\Lft,n=6}$ (i.e., the entanglement spectra) tend to share some notable similarities. Analogous reasonings apply to other pairs of channels as well. Because the effective dimension of the leading nontrivial block in $\tilde{\rho}_{\Lft,n}$ increases as $n$ approaches $\frac{N}{2}$, the saturation values of $S^{n}_{\textrm{vN}}$ and $S^{N-n}_{\textrm{vN}}$ get raised in a successive way. This is consistent with the overall tendency seen in Fig.~\ref{pic4}(d).

\subsection{Two types of the melting processes of the entanglement patterns in the thermal phase}

In this work, we exclusively focus on the even chains at the half filling with the periodic boundary conditions. Under these specifications, we find that there exist two types of the entanglement imbalance patterns depending on the parity of the total number of bosons. As shown by Figs.~\ref{pics1}(a),(b), when the total number of bosons is even, the discontinuous jumps of $S^n_{\textrm{vN}}$ occur only for those odd-$n$ channels. In comparison, as shown by Figs.~\ref{pics1}(c),(d), when the total number of bosons is odd, the discontinuous jumps of $S^n_{\textrm{vN}}$ occur only for those even-$n$ channels. Here we always begin with the initial $l$-state and choose the disorder strength $\mu=2J$ to be small.

To be pedantic, we shall name the entanglement imbalance pattern in Figs.~\ref{pics1}(a),(b) the reflection-symmetric one (with respect to the channel axis $n$). While, the entanglement imbalance pattern in Figs.~\ref{pics1}(c),(d) shall be called the reflection-asymmetric one for the obvious reason.

Due to the weak breaking of the spatial reflection symmetry at weak disorder, we know that these two types of entanglement imbalance patterns will melt in the long-time limit. Figure~\ref{pics1} reveals how the melting processes for these two entanglement patterns differ.

Concretely, from Figs.~\ref{pics1}(a),(b), we observe that for the reflection-symmetric entanglement pattern, the $\mathbb{Z}_2$ channel wave pattern with respect to the even versus odd channel indexes disappears but the channel reflection symmetry of the entanglement pattern persists. By contrast, from Figs.~\ref{pics1}(c),(d), we observe that for the reflection-asymmetric case, the disappearance of the $\mathbb{Z}_2$ channel wave is accompanied by an entanglement symmetrization process with the emergence or restoration of the channel reflection symmetry. Note that this channel reflection symmetry of the entanglement pattern does not exist in the initially generated reflection-asymmetric entanglement imbalance pattern. In this regard, the melting process of the reflection-asymmetric entanglement pattern is more dramatic than that of the reflection-symmetric one.

\subsection{Entropy inhomogeneity in the higher-energy, weak-disorder section}

Now we switch to the opposite extreme, the initial $p$-state, which maximizes the interaction. As displayed by Figs.~\ref{pic5}(a)-(c), in this case both the $\{S^{n}_{\textrm{vN}}\}$ evolution and the $\{p_{n,\infty}\}$ distribution alter drastically at small $\mu$.

First, once all the bosons are loaded onto a site, the Hubbard term dominates the Hamiltonian, which renders the single-boson tunnelings quenched as perturbations. To reduce the energy mismatch, the time evolution of the $p$-state tends to preserve its cluster structure. Further, the neighboring eigenstates available to the $p$-state also share the similar cluster features to sustain their comparable energy densities. Consequently, within this higher-energy interval, the translation and reflection symmetries of the model are bound to be broken in an abrupt way by the small $\mu$. Numerically, Fig.~\ref{pic5}(c) confirms the scaling trend of $p_{n,\infty}$ toward this interaction-enabled cluster localization at weak disorder upon increasing $L$.

Second, unlike the entanglement imbalance pattern formation in Figs.~\ref{pic4} and \ref{pics1}, starting from the $p$-state, $S^{n}_{\textrm{vN}}$ grows continuously from zero and no special pattern arises. Differing also from the long-time entropy symmetrization in Figs.~\ref{pic4} and \ref{pics1}, a strong entropy inhomogeneity develops inside the cluster MBL regime [see Figs.~\ref{pic5}(a),(b)]. A qualitative justification for this may run as follows. Still take the pair of channels $n=1,\ 6$ in an $L=14,~N=7$ chain with small $\mu$ as an example.
\begin{enumerate}
\item[(i)] According to Fig.~\ref{pic5}(c), the $n=1$ channel is dominated by the configurations with $6$ bosons moved to near the left entanglement cut and $1$ boson left within the left half-chain. While the $n=6$ channel is dominated by the configurations with $6$ bosons localized around the original leftmost site and $1$ boson hopping across the left cut into the right half-chain.
\item[(ii)] Because the leading energy mismatch between the initial $p$-state and the state in the channel $n=6$ is small, any additional fluctuations induced by the hoppings of single boson within the right half-chain are relatively important, thus these processes are restricted and the corresponding $S^{n=6}_{\textrm{vN}}$ gets suppressed.
\item[(iii)] In comparison, the leading energy mismatch between the initial $p$-state and the state in the channel $n=1$ is large, so comparatively, the fluctuations within the $n=1$ channel owing to the single-boson tunnelings along the left half-chain are less influential, suggesting that the corresponding hopping processes are more extended, i.e., the size of the pertinent block in $\tilde{\rho}_{\Lft,n=1}$ rises. Accordingly, after the normalization, $S^{n=1}_{\textrm{vN}}$ becomes enhanced.
\end{enumerate}
In this sense, it is the significant energy gap between the channels $n=1,\ 6$, together with its interplay with the weak disorder, that underpins the dynamics of the cluster MBL regime [see Figs.~\ref{pic5}(a)-(c)]. Other pairs of channels could be addressed in a similar way. Through manipulating the initial $p$-state, we therefore find the equilibrated coexistence of the particle and entropy inhomogeneities in one unified dynamical setting.

Finally, the above picture carries over to the strongly disordered circumstance with the addition that now each boson is localized by disorder as evident from Fig.~\ref{pic5}(f), thereby being confined to the spatial regions set by the localization length. This explains why the entropy inhomogeneity is reduced at $\mu=20J$ [see Figs.~\ref{pic5}(d),(e)]. Thus, this reduction is also consistent with the entropy symmetrization hypothesis proposed for the disorder-driven MBL phenomenology.         

\section{Summary and outlook} \label{sectionsummary}

We numerically explore the energy-resolved dynamical phase diagram of the 1D dBH model using the quantum quench evolution of the entanglement entropy based on the U(1) symmetry decomposition. The behavior of the Bose system at the lower energy densities is reminiscent of that of the fermionic (spin) model. Conversely, at the higher energy densities, a distinct cluster MBL regime emerges even at the weak disorder, characterized by the presence of neither an unbounded entropy growth nor a detrimental effect from the finite-size drift. In the lower-energy section of the phase diagram, despite the second law of thermodynamics preventing a decrease in the total entanglement entropy over time, there is a long-term reduction in the channel-resolved entropies for a thermalizing state. A detailed analysis of the entropy distribution amongst the individual channels also suggests that the previously observed double-logarithmic growth of the number entropy may not necessarily imply the eventual thermalization of the system. Rather, the double-logarithmic growth of the total number entropy as is also observed here does not actually contradict the localization phenomenon. By the inspection on an entanglement imbalance pattern generated from an initial product state, we hypothesize a universal entropy symmetrization process for the entire strongly disordered regime in the dynamical phase diagram of the dBH model.

In spite of the progress so far, several questions remain open regarding the analytical understanding of the descriptive framework for the cluster MBL regime \cite{ChenChenWang2025}, and the mechanism of the entanglement imbalance pattern generation as well as its melting. Protocols on how to experimentally measure the symmetry-resolved entanglement entropy comprise another promising direction for the future study. These continued efforts will undoubtedly reveal more surprises founded upon the interplay among quantum statistics, symmetry, entanglement, randomness, and interaction.

\section{Acknowledgements}

We thank Z.~Cai for the useful discussions. J.~C. and X.~W. were supported by MOST2022YFA1402701 and the NSFC Grant No.~11974244. C.~C. was supported by a start-up fund from Shanghai Jiao Tong University and the sponsorship from the Yangyang Development fund.

\appendix

\section{Finite-size scaling of the averaged level-spacing ratio} \label{app1}


It is crucial practice to critically study how the various phase-regime boundaries of the dynamical phase diagram of the 1D dBH model would change under the increase of the system size, although the numerically reached maximal chain length in this work remains quite limited. To this aim, we conduct in \cite{ChenChenWang2025} the extensive numerical calculations to obtain the four finite-size dynamical phase diagrams of the dBH chain at the four different chain lengths from $L=8$ up to $L=14$. The number of the loaded bosons in each size is always kept at the half-filling, namely $N=L/2$. The results are illustrated by Fig.~1 of \cite{ChenChenWang2025}. For the convenience of the comparison, we have intentionally chosen the scale of the color bar in each of these four panels of the figure to be the same.

There are two major pieces of information we can read out from this figure.
\begin{itemize}
\item First and foremost, there exists a pronounced drift of the phase-regime boundary between the thermal phase and the scatter MBL regime in the lower-energy section of the dynamical phase diagram. Such a drifting trend under the increase of $L$ toward the greater values of the disorder strength resembles what happens in the disordered Heisenberg $X\!X\!Z$ chain and is now well known to be the main obstacle toward reliably identifying MBL in the research community.
\item Second, by contrast, the cluster MBL regime in the higher-energy section of the dynamical phase diagram appears to be robust and stable. Especially, the phase-regime boundary between the cluster MBL regime and the thermal phase is nicely moving downward to around the normalized energy $\varepsilon\approx0.5$, suggesting the probable persistence of both the cluster MBL regime as well as the mobility edge in the large-size limit for the dBH chain. Please refer to Fig.~1 of \cite{ChenChenWang2025} for a detailed scaling analysis that bolsters this above claim.
\end{itemize}

Admittedly, the caveat drawn from the observed significant finite-size drift in Fig.~1 of Ref.~\cite{ChenChenWang2025} raises the critical concerns on the stability or the existence of the scatter MBL regime in the thermodynamic limit. There could be three possibilities. In the worst case, there would be no scatter MBL regime in the lower-energy section of the dynamical phase diagram of the dBH model when the chain length approaches infinity. The second possibility is that the scatter MBL regime does survive but its location is far beyond $\mu=20J$. The third possibility is the best that the yielded dynamical phase diagrams in Fig.~1 of Ref.~\cite{ChenChenWang2025} are qualitatively correct that when $\mu=20J$ the dBH chain would be genuinely fully localized even when taking the thermodynamic limit.

If the first or the second possibilities were selected by the infinite-$L$ dBH chain, then our reported results in Figs.~\ref{pic2_1}(a),(b), Fig.~\ref{pic3}(b), Fig.~\ref{pic4}(f), Fig.~\ref{pic9}(d), and the associated analyses or statements about them would all be no longer valid. Instead, under this circumstance, these presented localization results would be mostly replaced by those results showing the ubiquity of the thermal phase. Nevertheless, one of the central results of our work, namely, the entropy symmetrization hypothesis proposed in Sec.~\ref{sectionrltparttwo} for the strong-disorder section of the dynamical phase diagram of the dBH model stands still to be intact.

Apparently, at this stage all other findings regarding the thermal phase and the cluster MBL regime in the dynamical phase diagram of the dBH chain tend to be preserved against this above finite-size scaling analysis based on the disorder averaged level-spacing ratio \cite{Sierant_2018,mondragon2015many,sierant2017many}.

\section{Finite-size scaling of the averaged local boson density distribution} \label{app2}

\begin{figure*}[t]
\begin{center}
\includegraphics[width=1\linewidth]{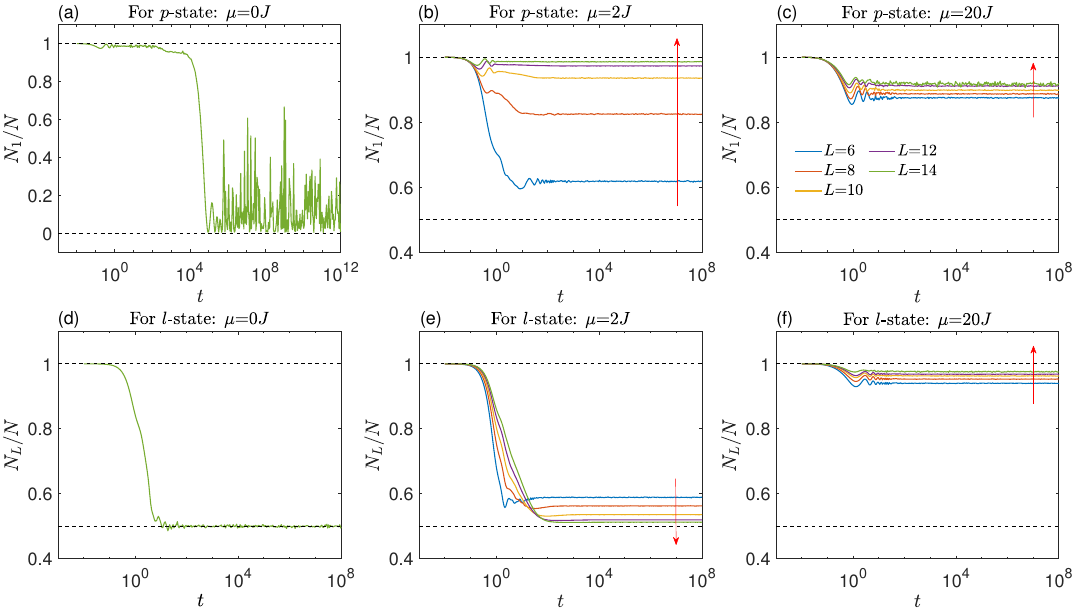}
\caption{The scaling of the quantum quench evolution of the local boson density distribution across {\color{black} the clean BH chain} and the four possible phase regimes in the dynamical phase diagram of the dBH chain. The initial state in the top row is set to be the higher-energy $p$-state with {\color{black} the disorder strength $\mu=0J$ in (a)}, $\mu=2J$ in (b), and $\mu=20J$ in (c). Here, $N_1$ stands for the total number of bosons on the first site of the periodic chain. The initial state in the bottom row is set to be the lower-energy $l$-state with {\color{black} the disorder strength $\mu=0J$ in (d)}, $\mu=2J$ in (e), and $\mu=20J$ in (f). Here, $N_L$ stands for the total number of bosons along the left half of the periodic chain. Notice that when increasing the chain length from $L=6$ to $14$, we always keep the system at the half-filling, i.e., $N=L/2$. Recall that $N$ is the total number of bosons in the clean BH or dBH chain. {\color{black} In view of the robust thermalization in (a),(d), to simplify the presentation, we only show the results at $L=14$ there.}}
\label{pic9}
\end{center}
\end{figure*}

\begin{figure*}[htb]
\begin{center}
\includegraphics[scale=0.97]{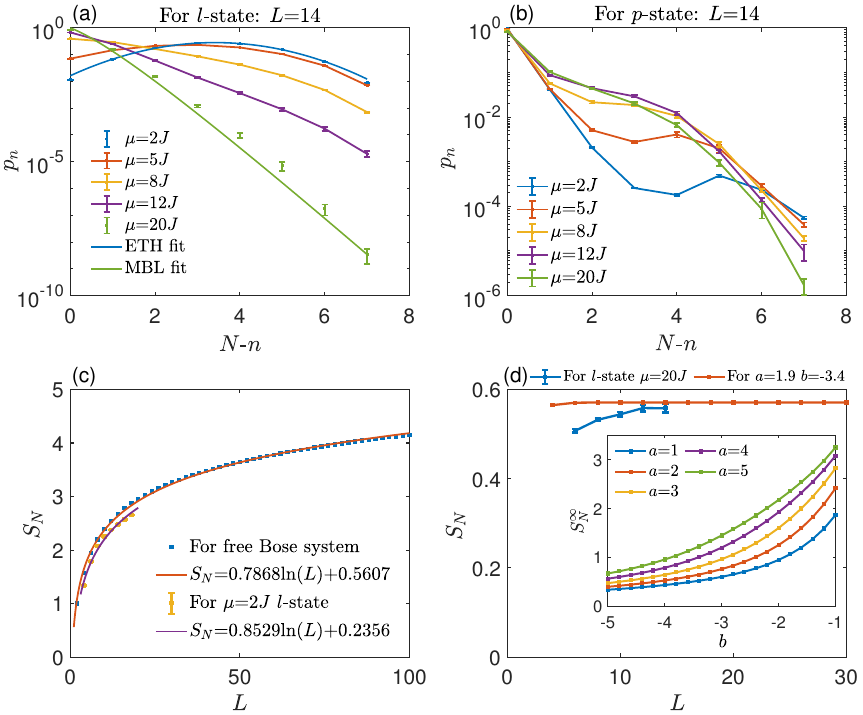}
\caption{(a),(b) The $p_n$ distribution of the long-time evolved state for a $L=14$ dBH chain with the different disorder strengths. (a) is for the initial $l$-state. Under the circumstance of $\mu=2J$, the $p_n$ distribution is fitted by a Gaussian distribution. For $\mu=20J$, the $p_n$ distribution is fitted by $p_{n_r}=\left( n_r+1 \right) ^ae^{bn_r+f}$ with $a=1.9$ and $b=-3.4$, where $n_r=N-n$ denotes the number of particles in the right half-chain. (b) is for the initial $p$-state. (c) The logarithmic fit of the particle number entropy $S_N$ as a function of the chain length for the long-time evolved state starting from the initial $l$-state and the disorder strength is set to be small. A comparison logarithmic fit to the ideal case of the free Bose system is also given. (d) The area law of the particle number entropy $S_N$ for the long-time evolved state when the initial state is the $l$-state and the disorder strength is large, and the inset gives the results of $S_N$ as a function of the parameters from the phenomenological analysis of an infinitely long chain. All the calculations related to the number entropy in the figure have taken $2$ as the logarithmic base.}
\label{pic10}
\end{center}
\end{figure*}

In the quantum quench dynamics, the generic distinction between the thermal phase and the putative localization regime is usually defined through whether the initial-state particle-density distribution would be surviving under the unitary time evolution of the Hamiltonian in the long-time and large-length limit or not \cite{SchreiberBloch,mondaini2017many}. For this purpose, in Fig.~\ref{pic9} of this appendix we carefully examine the scaling of the effectively infinite-time quantum quench results of the local boson density distribution. {\color{black} The first two panels with the disorder strength $\mu=0J$ of Fig.~\ref{pic9} are included for completeness to demonstrate the robust thermalization in the clean BH chain in both the lower- and the higher-energy sections of the dynamical phase diagram when the onsite Hubbard repulsion $U=3J$ is small. Accordingly, the Hilbert-space fragmentation \cite{SalaPollmann} expected to be pronounced in the disorder-free systems is absent and irrelevant to the present work. (The Hilbert-space fragmentation will play a role when the onsite Hubbard repulsion is extremely large.)} The remaining four panels of Fig.~\ref{pic9} correspond respectively to the identified four types of the phase regimes in the dynamical phase diagram of the dBH chain.
\begin{itemize}
{\color{black} \item[(a):] This top-left panel of Fig.~\ref{pic9} targets the higher-energy section of the dynamical phase diagram for the clean BH chain $(\mu=0J)$. Here, we start from the $p$-state: the initial occupancy on site $1$ is $N$, i.e., $N_1(t=0)/N=1$. As the time evolves, this initial occupation configuration persists for a long period, but eventually collapses to a steady thermal state as per the following two measures. First, the collapsed value of $N_1/N$ approaches $1/L$, consistent with the expectation for a thermal state. Second, assume thermalization, then for the $p$-state, it is a rough estimate that $U\frac{L}{2}(\frac{L}{2}-1)\approx UL\langle (\hat{n}-\frac{1}{2})^2\rangle$ where $\hat{n}$ is the local boson density operator and the expectation is taken upon the steady state. Clearly, $\langle (\hat{n}-\frac{1}{2})^2\rangle$ is the variance of the long-time value of $N_1$. As the energy encoded in the $p$-state is conserved and high, the fluctuation of the long-time evolution of $N_1/N$ is significant, and at the order of $1/\sqrt{L}$. This explains the large fluctuation of $N_1/N$ in this panel. (a) contrasts with (b).
\item[(d):] This bottom-left panel of Fig.~\ref{pic9} targets the lower-energy section of the dynamical phase diagram for the clean BH chain $(\mu=0J)$. We start from the $l$-state [see (\ref{lstate})] such that $N_L(t=0)/N=1$ where $N_L$ counts the total bosons on the left half-chain. As illustrated, this initial-state configuration is lost within ten seconds, and the clean chain promptly relaxes to a thermal phase where the expectation value of $N_L/N$ is $1/2$. Because the energy encoded in the $l$-state is low (almost a constant), the fluctuation of $N_1/N$ is at the order of $1/(L\sqrt{L})$. After smoothing over the left half-chain, the fluctuation of $N_L/N$ thus becomes negligible. (d) parallels (e).} 
\item[(b):] This top-middle panel of Fig.~\ref{pic9} targets the higher-energy, weak-disorder $(\mu=2J)$ section of the dynamical phase diagram, accommodating the cluster MBL regime. We start from the $p$-state with all the $N=L/2$ bosons loaded onto the first site, i.e., $N_1=N$ at $t=0$. Under the unitary time evolution, the boson concentration on the first site does decrease over time but eventually saturates to a quite significant value, demonstrating the breaking of the ergodicity. More importantly, with the increase of the chain length, the saturation value of the boson density on the first site for the initial $p$-state under the weak-disorder condition converges upward to $1$, showing the potential robustness of the localization occurring in this cluster MBL regime even in the thermodynamic limit. Notice that this trend toward localization from the scaling of the local boson density distribution is consistent with the trend seen in Fig.~1 of Ref.~\cite{ChenChenWang2025} from the scaling of the averaged level-spacing ratio.  
\item[(c):] The top-right panel of Fig.~\ref{pic9} switches to the higher-energy, strong-disorder $(\mu=20J)$ section of the dynamical phase diagram, focusing still on the cluster MBL regime. Start from the $p$-state with all the $N=L/2$ bosons put onto the first site, i.e., $N_1=N$ at $t=0$, the unitary time evolution initially reduces the boson concentration on the first site but eventually a considerable residue value of $N_1/N$ persists, signaling the disorder-assisted cluster localization sets in. Likewise, with the increase of the chain length, the saturation value of the boson density on the first site for the $p$-state under the strong-disorder condition also converges upward to $1$ but the tendency somehow slows down as compared to Fig.~\ref{pic9}(a). Taken together, Figs.~\ref{pic9}(a),(b) tend to suggest the overall stability of the cluster MBL regime in the thermodynamic limit. This trend also appears to be in line with the results of Fig.~1 of Ref.~\cite{ChenChenWang2025}.
\item[(e):] The bottom-middle panel of Fig.~\ref{pic9} demonstrates the normal relaxation of the dBH chain initialized from the $l$-state within the lower-energy window toward the thermal state when the disorder strength is weak $(\mu=2J)$. Note that here $N_L$ denotes the total number of bosons on the left half-chain. Under the increase of the system size, the long-time value of the ratio $N_L/N$ does converge downward to $1/2$, meaning that the $N=L/2$ bosons now have been uniformly distributed along the whole chain. The information about the initial $l$-state is then lost as expected.
\item[(f):] The bottom-right panel of Fig.~\ref{pic9} addresses the scatter MBL regime in the lower-energy, strong-disorder section of the dynamical phase diagram. Here, as before we begin with the $l$-state but set the disorder strength to be large, i.e., $\mu=20J$. While the scaling trend of the time evolution of $N_L/N$ in this section seems to be not inconsistent with the occurrence of the ergodicity breaking as it exhibits both the large deviation and the sharp contrast from that of Fig.~\ref{pic9}(c), in view of the continuous finite-size drift of the phase-regime boundary between the thermal phase and the scatter MBL regime as displayed by Fig.~1 of \cite{ChenChenWang2025} via the measure of the averaged level-spacing ratio, we cannot definitely tell the fate of the scatter MBL regime once approaching the long-time, large-size limit. This complication or dilemma seems to be absent in our examination of the cluster MBL regime.
\end{itemize}

\section{Analytics on the scaling behavior of the number entropy in the lower-energy section} \label{app3}

In this appendix, for the later reference, let
\begin{equation}
n_r=N-n
\end{equation}
denote the number of bosonic particles in the right half-chain, $n$ denote the number of bosonic particles in the left half-chain, and $N=L/2$ denote the total number of bosonic particles in the whole chain of length $L$.

The $l$-state, whose normalized energy is in the ETH phase, thermalizes at the small disorder strength. As shown by Fig.~\ref{pic10}(a), the corresponding long-time $p_{n_r}$ does approximately satisfy the Gaussian distribution when $\mu=2J$. [Recall that $p_{n_r}\ (p_n)$ represents the probability for the right (left) half-chain to contain the $N-n\ (n)$ bosons, so $p_{n_r}=p_n$.] One can thus assume that
\begin{equation}
p_{n_r}=\alpha \cdot e^{-\left( n_r-N/2 \right) ^2/\beta ^2}.
\end{equation}
Further, since $p_{n_r}$ needs to be normalized, as compared to the normalized Gaussian distribution, it is derivable that $\beta \sim N$ and $\alpha \sim \frac{1}{N}$. According to Fig.~\ref{pic4}(c) in the main text, it is then reasonable to let $\beta \approx cL$ and $\alpha \approx \frac{d}{L}$, where $c$, $d$ are the two parameters. Therefore, the particle number entropy $S_N$ can then be calculated as follows,
\begin{align}
S_N&=\sum_{n_r=0}^N{-p_{n_r}\cdot \log \left( p_{n_r} \right)} \nonumber \\
&=-\log \left( \alpha \right) +\frac{\alpha}{\beta ^2}2\sum_{n_r=0}^{N/2}{\left[ n_{r}^{2}e^{-n_{r}^{2}/\beta ^2} \right]} \nonumber \\
&\approx -\log \left( \alpha \right) +\frac{\alpha}{\beta ^2}2\int_0^{N/2}{n_{r}^{2}e^{-n_{r}^{2}/\beta ^2}dn_r} \nonumber \\
&=\log \left( L \right) -\log \left( d \right)+\frac{d}{4}\left[ 2c\sqrt{\pi}\textrm{Erf}\left( \frac{1}{4c} \right) -e^{-1/16c^2} \right]. \label{SNlogL}
\end{align}
We see that $S_N \sim \ln(L)$. This is also verified by the numerical results given in Fig.~\ref{pic10}(c).

For a free bosonic system without the interaction, an ideal case can be assumed where the various possible configurations are assumed to have the same probability in the thermalized state, which means that $p_n$ is determined only by the Hilbert-space dimension of the block matrix. The total Hilbert space of a bosonic system with the chain length $L$ and the number of particles $N$ is
\begin{equation}
D_{N}^{L}=\frac{\left( L+N-1 \right) !}{N!\left( L-1 \right) !},
\end{equation}
while the dimension of the Hilbert subspace with the particle number $n$ on the left half-chain is
\begin{equation}
D_n=D_{n}^{L/2}D_{N-n}^{L/2},
\end{equation}
so one finds that
\begin{equation}
p_n=D_n/D_{N}^{L}.
\end{equation}
For the current half-filled system, i.e., $N=L/2$, the result of the numerical calculation is shown in Fig.~\ref{pic10}(c), which also shows that $S_N \sim \ln(L)$.

For the case of the strong disorder, the system seems to be localized on a finite chain, as shown by Fig.~\ref{pic10}(a). In this case, $p_{n_r}$ can be roughly fitted with an exponential decay like function with
\begin{equation}
p_{n_r}=\left( n_r+1 \right) ^ae^{bn_r+f},
\end{equation}
where $b$ is always less than $0$ and the larger the disorder strength, the smaller $b$ will be. Using the normalization condition of $p_{n_r}$, we can obtain
\begin{equation}
p_{n_r}=\frac{\left( n_r+1 \right) ^ae^{b\left( n_r+1 \right)}}{\left[ \textrm{Li}_{-a}\left( e^b \right) -\left( e^b \right) ^{2+N}\textrm{L}\left( e^b,-a,2+N \right) \right]},
\end{equation}
where $\textrm{Li}_{-a}\left( e^b \right) $ is the Lerch zeta function, and $\textrm{L}\left( e^b,-a,2+N \right) $ is the polylogarithm function. In Fig.~\ref{pic10}(d), the calculation of $S_N$ using the normalized $p_{n_r}$ values is given, which can be seen to behave as per an area law. When the chain length is infinite, we can get
\begin{equation}
p_{n_r}=\left( n_r+1 \right) ^ae^{bn_r}\frac{e^b}{\textrm{Li}_{-a}\left( e^b \right)}.
\end{equation}
Taking this $p_{n_r}$ to compute $S_{N}^{\infty}$, turning the summation of the formula into an integral, a complex analytical expression can be obtained. In the inset of Fig.~\ref{pic10}(d), $S_{N}^{\infty}$ is calculated using the numerical integration, and it can be seen that as the disorder strength increases (i.e., $b$ decreases), $S_{N}^{\infty}$ becomes smaller, and the tendency for $S_{N}^{\infty}$ to become smaller slows down when the disorder is strong, which indicates that the eventual entropy does not decrease continuously to zero as the disorder strength increases.

\section{Additional details of the data fitting} \label{app4}

Figures~\ref{pic2} and \ref{pic2_1} of the main text illustrate the dynamical evolution of the particle-number entropy \( S_N \) and the configuration entropy \( S_C \) under varying normalized energy and disorder strength. The corresponding growth behaviors are summarized in Table~\ref{table1}. In this appendix, we systematically perform the quantitative fits to the time-dependent behaviors of \( S_N \) and \( S_C \) across the different phase regimes. For each case, we first specify the fitting time windows consistent with those used in the main text, and subsequently present the optimal fitting results along with the corresponding fitted parameter ranges.

\begin{figure*}[htb]
\begin{center}
\includegraphics[width=1\linewidth]{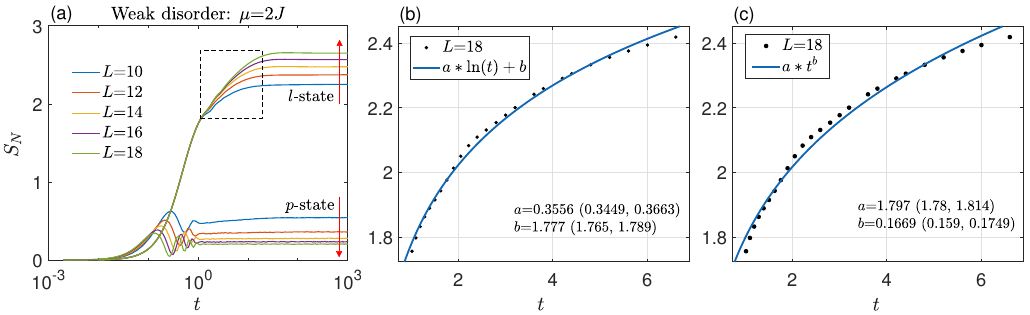}
\caption{(a) Time evolution of the particle-number entropy \( S_N(t) \) under the weak-disorder condition (\( \mu = 2J \)) for various system sizes \( L = 10, 12, 14, 16, 18 \). The black dashed box indicates the time window selected for data fitting. (b) Logarithmic fit of the data within the boxed region in (a) for \( L = 18 \) using the function \( S_N(t) = a \ln(t) + b \), yielding the best-fit parameters \( a = 0.3556\ (0.3449, 0.3663) \), \( b = 1.777\ (1.765, 1.789) \). (c) Power-law fit of the same set of data using \( S_N(t) = at^b \), with the fitted parameters \( a = 1.797\ (1.780, 1.814) \), \( b = 0.1669\ (0.159, 0.1749) \).}
\label{pic1s}
\end{center}
\end{figure*}

\begin{figure*}[htb]
\begin{center}
\includegraphics[width=0.7\linewidth]{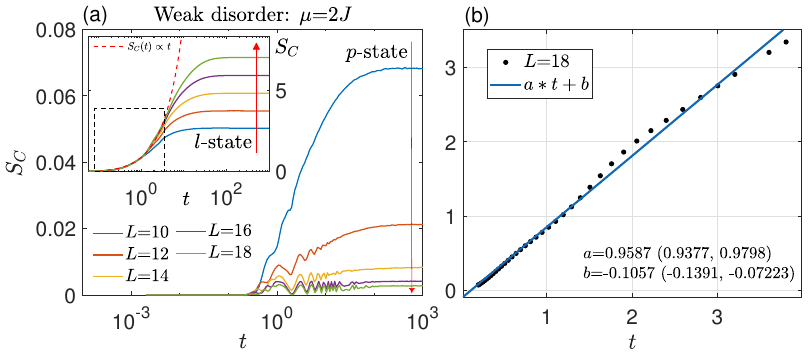}
\caption{(a) Time evolution of the configuration entropy \( S_C(t) \) under weak disorder \( (\mu = 2J) \) for various system sizes. The low-energy \( l \)-state and the high-energy \( p \)-state exhibit qualitatively distinct behaviors at long times. The black dashed box in the inset highlights the early-time window used for the linear fit shown in (b). (b) Linear fit of the early-time growth of \( S_C(t) \) for \( L = 18 \), using the form \( S_C(t) = a t + b \), with the best-fit parameters \( a = 0.9587\ (0.9377, 0.9798) \), \( b = -0.1057\ (-0.1391, -0.07223) \).}
\label{pic2s}
\end{center}
\end{figure*}  

{\it (i) Time evolution of \( S_N \) starting from the low-energy \( l \)-state at weak disorder.}---Figure~\ref{pic1s} presents the time evolution of the particle-number entropy \( S_N(t) \) under the weak disorder \( (\mu = 2J) \) for various system sizes. As shown in panel~(a), the entropy exhibits the rapid initial growth followed by a saturation at long times. To quantitatively characterize this intermediate-time dynamics, we focus on the data within the black dashed box for the largest system size \( L = 18 \) in Fig.~\ref{pic1s}. Panels~(b) and~(c) display the fitting results using two functional forms: a logarithmic function \( S_N(t) = a \ln(t) + b \) and a power-law function \( S_N(t) = a t^b \), respectively. A comparison between the two fitting models reveals that the logarithmic fit provides a more accurate representation of the data. This is evidenced by the visibly smaller deviation of the solid fitting line from the numerical data points in panel~(b), as well as the more stable and narrower confidence intervals for the fitted parameters.

\begin{figure*}[htb]
\begin{center}
\includegraphics[width=1\linewidth]{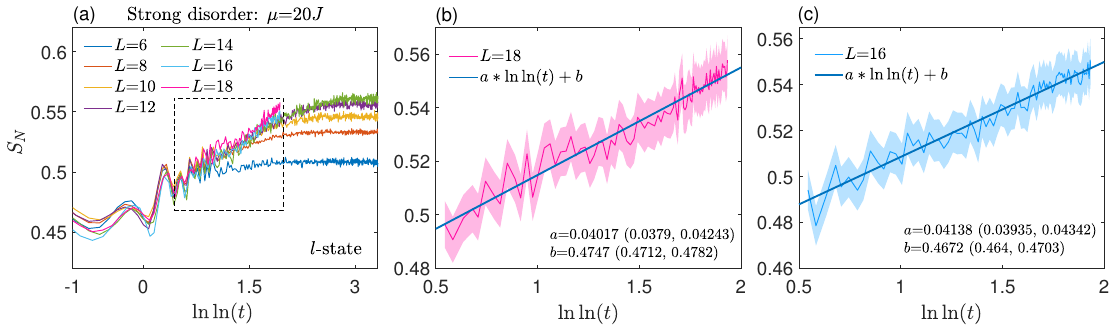}
\caption{(a) Time evolution of the particle-number entropy \( S_N(t) \) under strong disorder \( (\mu = 20J) \) for various system sizes, plotted against \( \ln \ln(t) \). The system is initialized in a low-energy \( l \)-state. The black dashed box highlights the time window used for fitting. (b) Double-logarithmic fit of the data for system size \( L = 18 \), using the form \( S_N(t) = a \ln \ln(t) + b \), yielding the best-fit parameters \( a = 0.04017\ (0.0379, 0.04243) \), \( b = 0.4747\ (0.4712, 0.4782) \). (c) The same fitting procedure applied to \( L = 16 \) yields the best-fit parameters \( a = 0.04138\ (0.03935, 0.04342) \), \( b = 0.4672\ (0.4644, 0.4703) \). Shaded regions are error bars.}
\label{pic3s}
\end{center}
\end{figure*}

\begin{figure*}[h]
\begin{center}
\includegraphics[scale=0.95]{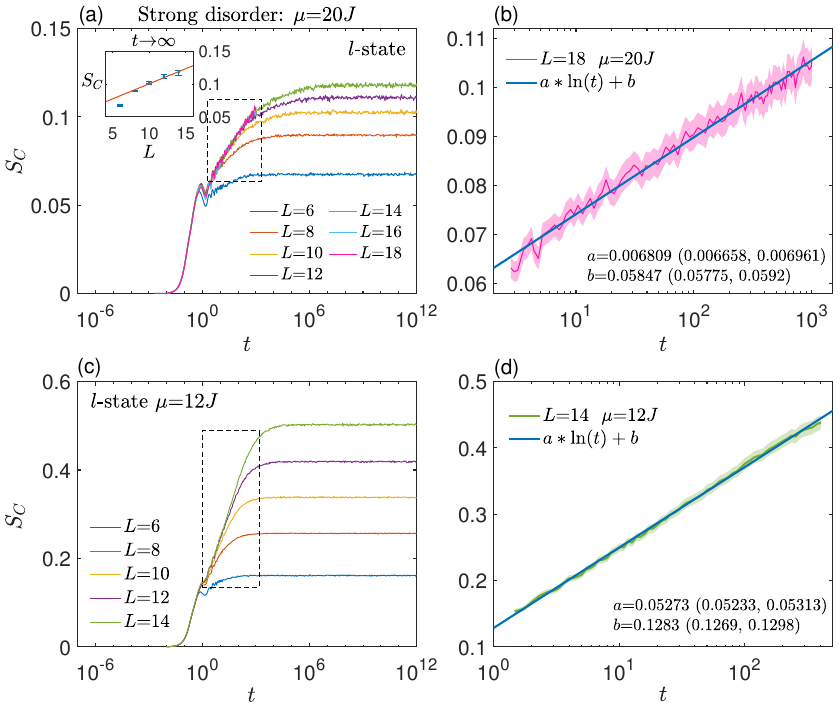}
\caption{(a) Time evolution of the configuration entropy $S_C$ for different system sizes $L$ in the presence of the strong disorder $(\mu=20J)$. The system is initialized in an $l$-state, which corresponds to a low-energy configuration. The black dashed box highlights the region used for fitting. The fitting is detailed in panel (b). The inset shows the long-time saturated value $S_C(t\to\infty)$ versus $L$, indicating a linear dependence. (c) Time evolution of $S_C$ for a different disorder strength $(\mu=12J)$, also starting from an $l$-state. (d) The logarithmic time dependence of $S_C$ is revealed for $L=14$ and $\mu=12J$, focusing on the region enclosed by the black dashed box in (c). The blue solid line in (b) and (d) represents a log fit to the data as per $S_C = a \ln(t) + b$, with the fitted parameters and their confidence intervals displayed. Shaded regions are error bars.}
\label{pic4s}
\end{center}
\end{figure*}

\begin{figure*}[h]
\begin{center}
\includegraphics[width=1\linewidth]{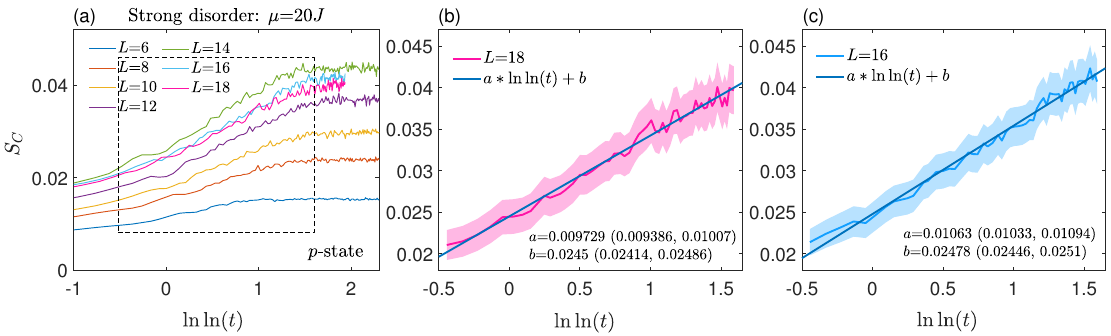}
\caption{(a) Time evolution of the configuration entropy \( S_C \) plotted as a function of \( \ln \ln(t) \) for various system sizes \( L \) under strong disorder \( (\mu = 20J) \). The system is initialized in a high-energy \( p \)-state. The black dashed box highlights the time window selected for further analysis. (b) Double-log growth of \( S_C \) as a function of \( \ln \ln(t) \) for system size \( L = 18 \), targeting the fitting region indicated in (a). The blue solid line denotes a linear fit of the form \( S_C(t) = a \ln\ln(t) + b \), with the extracted fitting parameters and their confidence intervals provided. (c) Double-log time dependence and the corresponding linear fit for the system size \( L = 16 \), following the same fitting procedure as in (b). The shaded regions in (b),(c) quantify the error bars.}
\label{pic5s}
\end{center}
\end{figure*}

To quantitatively evaluate the quality of the two fitting models, we compare several standard statistical indicators derived from the respective fitting procedures. For the logarithmic fit \( S_N(t) = a \ln(t) + b \), we obtain a coefficient of determination \( R^2 = 0.99519 \), an adjusted coefficient of determination \( R^2_{\mathrm{adj}} = 0.99498 \), a sum of squared errors (SSE) of 0.0047638, and a root mean square error (RMSE) of 0.014392, with degrees of freedom for error (DFE) equal to 23. In contrast, the power-law fit \( S_N(t) = a t^b \) yields a lower \( R^2 = 0.98807 \), \( R^2_{\mathrm{adj}} = 0.98755 \), a higher SSE of 0.011819, and a larger RMSE of 0.022669, under the same DFE. These statistical metrics indicate that the logarithmic model provides a better fit to the data. The higher values of \( R^2 \) and \( R^2_{\mathrm{adj}} \) demonstrate that the logarithmic form captures a greater proportion of the variance in the entropy growth. Simultaneously, the lower SSE and RMSE values confirm the improved predictive accuracy and the smaller residual deviations. Taken together, with the agreement shown in Fig.~\ref{pic1s}(b), these results support the claim that the intermediate-time dynamics of \( S_N(t) \) are more accurately described by a logarithmic growth than by a power-law scaling behavior \cite{Maximilian2020,kiefer2021slow,KieferEmmanouilidisPRL}.

{\it (ii) Time evolution of \( S_C \) starting from the low-energy \( l \)-state at weak disorder.}---We continue to focus on the thermal phase where it is known that the configuration entropy \( S_C(t) \) grows rapidly. For convenience, we repeat Fig.~\ref{pic2}(b) in Fig.~\ref{pic2s}, which illustrates the time evolution of the configuration entropy \( S_C(t) \) at weak disorder \( (\mu = 2J) \) across a range of system sizes. In panel~(a), we observe that the long-time saturation values of \( S_C(t) \) differ significantly between the low-energy \( l \)-state and the high-energy \( p \)-state, indicating qualitatively distinct dynamical behaviors. To characterize in detail the time evolution, we perform the fit within the black dashed box in the inset: a linear fit to the short early-time regime. Beyond that, there is a prolonged late-time relaxation regime. This relaxation regime is not captured by the linear-$t$ fit. Admittedly, although we use the Krylov method, due to the high local occupancy in Bose systems, the system sizes reached remain limited, so considering the narrowness of this early-time window, we intentionally plot the data in a log scale to demonstrate more clearly the enlargement of this early-time regime under the increase of system size.

Figure~\ref{pic2s}(b) shows that the initial growth of \( S_C(t) \) is described by a linear function \( S_C(t) = a t + b \), suggesting a near-ballistic initial spreading of the entanglement in the configurational space. The quality of the fit and the stability of the parameters support the claim that the configuration entropy exhibits a two-stage dynamical structure: an initial linear growth followed by a saturating relaxation. This result is consistent with the behavior expected for the weakly disordered systems in the thermal phase. Particularly, Kim and Huse observed a similar two-stage entanglement growth in Fig.~1 of their seminal paper \cite{KimHuse}, reporting the ballistic growth of the entanglement entropy in a nonintegrable ergodic spin chain.

{\it (iii) Time evolution of \( S_N \) starting from the low-energy \( l \)-state at strong disorder.}---Figure~\ref{pic3s} presents the time evolution of the particle-number entropy \( S_N(t) \) in the strong-disorder regime \( (\mu = 20J) \), with the system initialized in a low-energy \( l \)-state. As shown in panel~(a), the entropy exhibits extremely slow and nontrivial growth across a range of system sizes, indicative of glassy and highly localized dynamics. When plotted against \( \ln \ln(t) \), the data collapse onto an approximately linear line in the intermediate-time window before saturation, as indicated by the black dashed box.

Panels~(b) and~(c) show the fits for system sizes \( L = 18 \) and \( L = 16 \), respectively, using the functional form \( S_N(t) = a \ln \ln(t) + b \). The excellent agreement between the numerical data and the fit, along with the consistency of the fitting parameters, suggests a robust \( \ln \ln(t) \) scaling of the entropy growth. This double-logarithmic behavior is significantly slower than the logarithmic growth observed in weakly disordered, thermalizing systems, and can be interpreted as related to MBL or MBL-like phenomenology.

From a theoretical standpoint, \( \ln \ln(t) \) scaling may be attributed to the rare-region Griffiths effects \cite{KieferEmmanouilidisPRL,aceituno2024ultraslow} or to the strong disorder renormalization group fixed points \cite{Vosk2013many}, where entanglement dynamics and information spreading are governed by extremely slow and broadly distributed timescales. In such regimes, particle transport is effectively frozen, and only rare resonant spots contribute to the weak, yet nonzero, entropy growth. The observed behavior thus reflects the ultraslow dynamics of a system initialized in a low-energy \( l \)-state under strong disorder, consistent with the asymptotic dynamical localization and the absence of thermalization in the strong-disorder limit.

{\it (iv) Time evolution of \( S_C \) starting from the low-energy \( l \)-state at strong disorder.}---Figure~\ref{pic4s} illustrates the dynamical behavior of the configuration entropy \( S_C \) in a disordered system, highlighting its time evolution and the dependence on the system size \( L \). Panels~(a) and~(c) show the growth and the eventual saturation of \( S_C \) for the two disorder strengths \( \mu = 20J \) and \( \mu = 12J \), respectively, with the system initialized in a low-energy \( l \)-state. In both cases, \( S_C(t) \) exhibits an initial increase followed by a saturation at long times; the saturated value of the configuration entropy exhibits a volume-law scaling with the system size, indicating that the number of accessible configurations continues to extensively grow. As a consequence, the total entanglement entropy displays the unbounded growth even within the MBL regime \cite{BardarsonPollmannMoore,Serbyn2013}.

Panels~(b) and~(d) provide a detailed analysis of the intermediate-time regime, as indicated by the black dashed boxes in panels~(a) and~(c). In this temporal window, the entropy exhibits a clear logarithmic growth, well described by the functional form \( S_C(t) = a \ln(t) + b \). This behavior is characteristic of the slow information transport processes in disordered systems, where entropy grows slowly due to the hindered configurational exploration. The results of the fitting, including the extracted parameters \( a \) (the growth rate) and \( b \) (the initial offset), are reported along with their confidence intervals, confirming the accuracy and stability of the logarithmic description. Notably, the extracted slope \( a \) in panel~(d), corresponding to a weaker disorder strength \( (\mu = 12J) \), is larger than that in panel~(b), indicating a faster entropy growth rate as the disorder strength is smaller.

{\it (v) Time evolution of \( S_C \) starting from the high-energy \( p \)-state at strong disorder.}---First of all, notice that the scale of $S_C$ in Fig.~\ref{pic2_1}(d) is small. After considering the error bars, one perceives that the overall trend of the entropy-growth curves is tentatively aligned with each other. Moreover, once the system size is above $L=14$, this converging trend becomes more visible, suggesting that the finite-size effects may start to reduce beyond this size. Below, we explicitly perform the data analyses to show that the time development of $S_C$ in this circumstance follows a double-log scaling law.

Figure~\ref{pic5s}(a) repeats Fig.~\ref{pic2_1}(d), and presents the temporal evolution of the configuration entropy \( S_C \) in a strongly disordered system \( (\mu = 20J) \), with the system initialized in a high-energy \( p \)-state. Figure~\ref{pic5s}(a) shows \( S_C \) as a function of \( \ln\ln(t) \) for various system sizes \( L \), revealing a distinct intermediate-time growth regime. The black dashed rectangle there indicates the temporal window selected for the further analysis of the entropy dynamics.

Figures~\ref{pic5s}(b) and~(c) provide a closer examination of this intermediate-time behavior by plotting \( S_C \) versus \( \ln\ln(t) \) for system sizes \( L = 18 \) and \( L = 16 \), respectively. Within the identified fitting window, a clear linear dependence of \( S_C \) upon \( \ln\ln(t) \) is observed, consistent with a double-logarithmic growth law of the form \( S_C(t) = a \ln\ln(t) + b \). This behavior is indicative of the ultraslow dynamics in the presence of the strong disorder. The blue solid lines in Figs.~\ref{pic5s}(b) and~(c) represent the linear fits, and the extracted parameters—slope \( a \) and intercept \( b \)—are then reported alongside their respective confidence intervals.

The positive values of \( a \) confirm the monotonic increase of the configurational entropy within this time regime. The observed slight decrease in the slope \( a \) with the increasing system size hints at the possibility that, in the thermodynamic limit, the configurational entropy \( S_C \) may seldom exhibit visible growth (as the slope is too small but nonzero) when the system is initialized in a high-energy \( p \)-state.

\bibliography{disBH}

\end{document}